\documentclass[twocolumn, superscriptaddress]{revtex4-1} \usepackage{amsmath} \usepackage{booktabs}
\usepackage[utf8]{inputenc} \usepackage{xcolor} \usepackage{graphicx} \usepackage{hyperref}
\usepackage{orcidlink}
\usepackage{lineno}
\usepackage{acronym}

\renewcommand{\eqref}[1]{Eq.~(\ref{#1})}

\newcommand{\figref}[1]{Fig.~\ref{#1}}
\newcommand{\secref}[1]{Sec.~\ref{#1}}

\begin{document}

\title{Method to get Better Sky Maps in a GstLAL Low-Latency Analysis}

\author{Prathamesh Joshi \orcidlink{0000-0002-4148-4932}}
\email{prathamesh.joshi@ligo.org}
\affiliation{Department of Physics, The Pennsylvania State University, University Park, PA 16802, USA}
\affiliation{Institute for Gravitation and the Cosmos, The Pennsylvania State University, University Park, PA 16802, USA}
\affiliation{School of Physics, Georgia Institute of Technology, Atlanta, GA 30332, USA}

\author{Becca Ewing}
%\email{rebecca.ewing@ligo.org}
\affiliation{Department of Physics, The Pennsylvania State University, University Park, PA 16802, USA}
\affiliation{Institute for Gravitation and the Cosmos, The Pennsylvania State University, University Park, PA 16802, USA}

\author{Chad Hanna}
%\email{chad.hanna@ligo.org}
\affiliation{Department of Physics, The Pennsylvania State University, University Park, PA 16802, USA}
\affiliation{Institute for Gravitation and the Cosmos, The Pennsylvania State University, University Park, PA 16802, USA}
\affiliation{Department of Astronomy and Astrophysics, The Pennsylvania State University, University Park, PA 16802, USA}
\affiliation{Institute for Computational and Data Sciences, The Pennsylvania State University, University Park, PA 16802, USA}

\author{Zach Yarbrough \orcidlink{0000-0002-9825-1136}}
%\email{zach.yarbrough@ligo.org}
\affiliation{Department of Physics and Astronomy, Louisiana State University, Baton Rouge, LA 70803, USA}

\author{Jolien D. E. Creighton \orcidlink{0000-0003-3600-2406}}
%\email{jolien.creighton@ligo.org}
\affiliation{Leonard E.\ Parker Center for Gravitation, Cosmology, and Astrophysics, University of Wisconsin-Milwaukee, Milwaukee, WI 53201, USA}

\author{Shomik Adhicary \orcidlink{0009-0004-2101-5428}}
%\email{shomik.adhicary@ligo.org}
\affiliation{Department of Physics, The Pennsylvania State University, University Park, PA 16802, USA}
\affiliation{Institute for Gravitation and the Cosmos, The Pennsylvania State University, University Park, PA 16802, USA}

\author{Pratyusava Baral \orcidlink{0000-0001-6308-211X}}
%\email{pratyusava.baral@ligo.org}
\affiliation{Leonard E.\ Parker Center for Gravitation, Cosmology, and Astrophysics, University of Wisconsin-Milwaukee, Milwaukee, WI 53201, USA}

\author{Amanda Baylor \orcidlink{0000-0003-0918-0864}}
%\email{amanda.baylor@ligo.org}
\affiliation{Leonard E.\ Parker Center for Gravitation, Cosmology, and Astrophysics, University of Wisconsin-Milwaukee, Milwaukee, WI 53201, USA}

\author{Kipp Cannon \orcidlink{0000-0003-4068-6572}}
%\email{kipp.cannon@ligo.org}
\affiliation{RESCEU, The University of Tokyo, Tokyo, 113-0033, Japan}

\author{Sarah Caudill}
%\email{sarah.caudill@ligo.org}
\affiliation{Department of Physics, University of Massachusetts, Dartmouth, MA 02747, USA}
\affiliation{Center for Scientific Computing and Data Science Research, University of Massachusetts, Dartmouth, MA 02747, USA}

\author{Michael W. Coughlin \orcidlink{0000-0002-8262-2924}}
\affiliation{School of Physics and Astronomy, University of Minnesota, Minneapolis, Minnesota 55455, USA}

\author{Bryce Cousins \orcidlink{0000-0002-7026-1340}}
%\email{bryce.cousins@ligo.org}
\affiliation{Department of Physics, University of Illinois, Urbana, IL 61801 USA}
\affiliation{Department of Physics, The Pennsylvania State University, University Park, PA 16802, USA}
\affiliation{Institute for Gravitation and the Cosmos, The Pennsylvania State University, University Park, PA 16802, USA}

\author{Heather Fong}
%\email{heather.fong@ligo.org}
\affiliation{Department of Physics and Astronomy, University of British Columbia, Vancouver, BC, V6T 1Z4, Canada}
\affiliation{RESCEU, The University of Tokyo, Tokyo, 113-0033, Japan}
\affiliation{Graduate School of Science, The University of Tokyo, Tokyo 113-0033, Japan}

\author{Richard N. George \orcidlink{0000-0002-7797-7683}}
%\email{richard.george@ligo.org}
\affiliation{Center for Gravitational Physics, University of Texas at Austin, Austin, TX 78712, USA}

\author{Shaon Ghosh}
\affiliation{Montclair State University, 1 Normal Ave, Montclair, NJ 07042}

\author{Patrick Godwin \orcidlink{0000-0002-7489-4751}}
%\email{patrick.godwin@ligo.org}
\affiliation{LIGO Laboratory, California Institute of Technology, MS 100-36, Pasadena, California 91125, USA}
\affiliation{Department of Physics, The Pennsylvania State University, University Park, PA 16802, USA}
\affiliation{Institute for Gravitation and the Cosmos, The Pennsylvania State University, University Park, PA 16802, USA}

\author{Reiko Harada}
%\email{reiko.harada@ligo.org}
\affiliation{RESCEU, The University of Tokyo, Tokyo, 113-0033, Japan}
\affiliation{Graduate School of Science, The University of Tokyo, Tokyo 113-0033, Japan}

\author{Yun-Jing Huang \orcidlink{0000-0002-2952-8429}}
%\email{yun-jing.huang@ligo.org}
\affiliation{Department of Physics, The Pennsylvania State University, University Park, PA 16802, USA}
\affiliation{Institute for Gravitation and the Cosmos, The Pennsylvania State University, University Park, PA 16802, USA}

\author{Rachael Huxford}
%\email{huxfo013@umn.edu}
\affiliation{Minnesota Supercomputing Institute, University of Minnesota, Minneapolis, MN 55455, USA}

\author{James Kennington \orcidlink{0000-0002-6899-3833}}
%\email{james.kennington@ligo.org}
\affiliation{Department of Physics, The Pennsylvania State University, University Park, PA 16802, USA}
\affiliation{Institute for Gravitation and the Cosmos, The Pennsylvania State University, University Park, PA 16802, USA}

\author{Soichiro Kuwahara}
%\email{soichiro.kuwahara@ligo.org}
\affiliation{RESCEU, The University of Tokyo, Tokyo, 113-0033, Japan}
\affiliation{Graduate School of Science, The University of Tokyo, Tokyo 113-0033, Japan}

\author{Alvin K. Y. Li \orcidlink{0000-0001-6728-6523}}
%\email{alvin.li@ligo.org}
\affiliation{RESCEU, The University of Tokyo, Tokyo, 113-0033, Japan}
\affiliation{Graduate School of Science, The University of Tokyo, Tokyo 113-0033, Japan}

\author{Ryan Magee \orcidlink{0000-0001-9769-531X}}
%\email{ryan.magee@ligo.org}
\affiliation{LIGO Laboratory, California Institute of Technology, Pasadena, CA 91125, USA}

\author{Duncan Meacher \orcidlink{0000-0001-5882-0368}}
%\email{duncan.meacher@ligo.org}
\affiliation{Leonard E.\ Parker Center for Gravitation, Cosmology, and Astrophysics, University of Wisconsin-Milwaukee, Milwaukee, WI 53201, USA}

\author{Cody Messick \orcidlink{0000-0002-8230-3309}}
%\email{cody.messick@ligo.org}
\affiliation{Leonard E.\ Parker Center for Gravitation, Cosmology, and Astrophysics, University of Wisconsin-Milwaukee, Milwaukee, WI 53201, USA}

\author{Soichiro Morisaki \orcidlink{0000-0002-8445-6747}}
%\email{soichiro.morisaki@ligo.org}
\affiliation{Institute for Cosmic Ray Research, The University of Tokyo, 5-1-5 Kashiwanoha, Kashiwa, Chiba 277-8582, Japan}

\author{Debnandini Mukherjee  \orcidlink{0000-0001-7335-9418}}
%\email{debnandini.mukherjee@ligo.org}
\affiliation{NASA Marshall Space Flight Center, Huntsville, AL 35811, USA}
\affiliation{Center for Space Plasma and Aeronomic Research, University of Alabama in Huntsville, Huntsville, AL 35899, USA}

\author{Wanting Niu \orcidlink{0000-0003-1470-532X}}
%\email{wanting.niu@ligo.org}
\affiliation{Department of Physics, The Pennsylvania State University, University Park, PA 16802, USA}
\affiliation{Institute for Gravitation and the Cosmos, The Pennsylvania State University, University Park, PA 16802, USA}

\author{Alexander Pace \orcidlink{0009-0003-4044-0334}}
%\email{alexander.pace@ligo.org}
\affiliation{Department of Physics, The Pennsylvania State University, University Park, PA 16802, USA}
\affiliation{Institute for Gravitation and the Cosmos, The Pennsylvania State University, University Park, PA 16802, USA}

\author{Cort Posnansky \orcidlink{0009-0009-7137-9795}}
%\email{cort.posnansky@ligo.org}
\affiliation{Department of Physics, The Pennsylvania State University, University Park, PA 16802, USA}
\affiliation{Institute for Gravitation and the Cosmos, The Pennsylvania State University, University Park, PA 16802, USA}

\author{Anarya Ray \orcidlink{0000-0002-7322-4748}}
%\email{anarya.ray@ligo.org}
\affiliation{Leonard E.\ Parker Center for Gravitation, Cosmology, and Astrophysics, University of Wisconsin-Milwaukee, Milwaukee, WI 53201, USA}
\affiliation{Center for Interdischiplinary Exploration and Research in Astrophysics, Northwestern University, IL 60201, USA}

\author{Surabhi Sachdev \orcidlink{0000-0002-0525-2317}}
%\email{surabhi.sachdev@ligo.org}
\affiliation{School of Physics, Georgia Institute of Technology, Atlanta, GA 30332, USA}
\affiliation{Leonard E.\ Parker Center for Gravitation, Cosmology, and Astrophysics, University of Wisconsin-Milwaukee, Milwaukee, WI 53201, USA}

\author{Shio Sakon \orcidlink{0000-0002-5861-3024}}
%\email{shio.sakon@ligo.org}
\affiliation{Department of Physics, The Pennsylvania State University, University Park, PA 16802, USA}
\affiliation{Institute for Gravitation and the Cosmos, The Pennsylvania State University, University Park, PA 16802, USA}

\author{Stefano Schmidt \orcidlink{0000-0002-8206-8089}}
%\email{s.schmidt@uu.nl}
\affiliation{Nikhef, Science Park 105, 1098 XG, Amsterdam, The Netherlands}
\affiliation{Institute for Gravitational and Subatomic Physics (GRASP), Utrecht University, Princetonplein 1, 3584 CC Utrecht, The Netherlands}

\author{Urja Shah \orcidlink{0000-0001-8249-7425}}
%\email{urja.shah@ligo.org}
\affiliation{School of Physics, Georgia Institute of Technology, Atlanta, GA 30332, USA}

\author{Divya Singh \orcidlink{0000-0001-9675-4584}}
%\email{divya.singh@ligo.org}
\affiliation{Department of Physics, The Pennsylvania State University, University Park, PA 16802, USA}
\affiliation{Institute for Gravitation and the Cosmos, The Pennsylvania State University, University Park, PA 16802, USA}
\affiliation{Department of Physics, University of California, Berkeley, CA 94720, USA}

\author{Ron Tapia}
%\email{ron.tapia@ligo.org}
\affiliation{Department of Physics, The Pennsylvania State University, University Park, PA 16802, USA}
\affiliation{Institute for Computational and Data Sciences, The Pennsylvania State University, University Park, PA 16802, USA}

\author{Leo Tsukada  \orcidlink{0000-0003-0596-5648}}
%\email{leo.tsukada@ligo.org}
\affiliation{Department of Physics, The Pennsylvania State University, University Park, PA 16802, USA}
\affiliation{Institute for Gravitation and the Cosmos, The Pennsylvania State University, University Park, PA 16802, USA}
\affiliation{Department of Physics and Astronomy, University of Nevada, Las Vegas, 4505 South Maryland Parkway, Las Vegas, NV 89154, USA}
\affiliation{Nevada Center for Astrophysics, University of Nevada, Las Vegas, NV 89154, USA}

\author{Koh Ueno \orcidlink{0000-0003-3227-6055}}
%\email{koh.ueno@ligo.org}
\affiliation{RESCEU, The University of Tokyo, Tokyo, 113-0033, Japan}

\author{Aaron Viets \orcidlink{0000-0002-4241-1428}}
%\email{aaron.viets@ligo.org}
\affiliation{Concordia University Wisconsin, Mequon, WI 53097, USA}

\author{Leslie Wade}
%\email{leslie.wade@ligo.org}
\affiliation{Department of Physics, Hayes Hall, Kenyon College, Gambier, Ohio 43022, USA}

\author{Madeline Wade \orcidlink{0000-0002-5703-4469}}
%\email{madeline.wade@ligo.org}
\affiliation{Department of Physics, Hayes Hall, Kenyon College, Gambier, Ohio 43022, USA}

\author{Noah Zhang \orcidlink{0009-0003-3361-5538}}
%\email{noah.zhang@ligo.org}
\affiliation{School of Physics, Georgia Institute of Technology, Atlanta, GA 30332, USA}

\date{\today}

\keywords{Suggested keywords} %Use showkeys class option if keyword display desired

\begin{abstract}
Modeled gravitational wave 
searches correlate the strain data with a bank of gravitational wave template waveforms to
make detections of gravitational wave candidates, and these results are processed by
downstream tools to calculate the likely sky location and distance of the source of the candidates.
This is crucial for multi-messenger efforts, since it informs astronomers where to point their
telescopes to facilitate electromagnetic follow-up of the gravitational wave candidates.
We present a novel method to improve the low-latency results of the GstLAL gravitational wave
search pipeline, and thus improving sky location estimates of low-latency candidates.
This method involves ingesting the GstLAL low-latency results, and performing
a small targeted hierarchical search to recover the candidates with more accurate parameters,
in a medium-latency timescale (few seconds to five minutes). To test our method, we perform
a GstLAL low-latency analysis on forty days of data from the third observing run of LIGO,
Virgo, and KAGRA, and show that our method improves the GstLAL results by 5.38\% and the
subsequent sky location results by 16.75\% on average. In addition to this increase in precision,
we also show that these results are more accurate as compared to the GstLAL results.
This method has been adopted by GstLAL for the fourth observing run.

\end{abstract}

\maketitle

% ======================
%  ACRONYMS
% ======================
\acrodef{LSC}[LSC]{LIGO Scientific Collaboration}
\acrodef{LVC}[LVC]{LIGO Scientific and Virgo Collaboration}
\acrodef{LVK}[LVK]{LIGO Scientific, Virgo and KAGRA Collaboration}
\acrodef{aLIGO}{Advanced Laser Interferometer Gravitational-Wave Observatory}
\acrodef{aVirgo}{Advanced Virgo}
\acrodef{LIGO}[LIGO]{Laser Interferometer Gravitational-Wave Observatory}
\acrodef{IFO}[IFO]{interferometer}
\acrodef{LHO}[LHO]{LIGO-Hanford}
\acrodef{LLO}[LLO]{LIGO-Livingston}
\acrodef{O2}[O2]{second observing run}
\acrodef{O1}[O1]{first observing run}
\acrodef{O3}[O3]{third observing run}
\acrodef{O3a}[O3a]{first half of the third observing run}
\acrodef{O3b}[O3b]{second half of the third observing run}
\acrodef{O4a}[O4a]{first part of the fourth observing run}
\acrodef{O4}[O4]{fourth observing run}

\acrodef{SSM}[SSM]{subsolar-mass}
\acrodef{BH}[BH]{black hole}
\acrodef{BBH}[BBH]{binary black hole}
\acrodef{BNS}[BNS]{binary neutron star}
\acrodef{IMBH}[IMBH]{intermediate-mass black hole}
\acrodef{NS}[NS]{neutron star}
\acrodef{BHNS}[BHNS]{black hole--neutron star binaries}
\acrodef{NSBH}[NSBH]{neutron star--black hole binary}
\acrodef{PBH}[PBH]{primordial black hole binaries}
\acrodef{CBC}[CBC]{compact binary coalescence}
\acrodef{GW}[GW]{gravitational wave}
\acrodef{GWH}[GW]{gravitational-wave}
\acrodef{DBH}[DBH]{dispasstive black hole binaries}

\acrodef{SNR}[SNR]{signal-to-noise ratio}
\acrodef{FAR}[FAR]{false alarm rate}
\acrodef{PSD}[PSD]{power spectral density}
\acrodef{GR}[GR]{general relativity}
\acrodef{NR}[NR]{numerical relativity}
\acrodef{PN}[PN]{post-Newtonian}
\acrodef{EOB}[EOB]{effective-one-body}
\acrodef{ROM}[ROM]{reduced-order model}
\acrodef{IMR}[IMR]{inspiral--merger--ringdown}
\acrodef{EOS}[EoS]{equation of state}
\acrodef{FF}[FF]{fitting factor}
\acrodef{FT}[FT]{Fourier Transform}

\acrodef{LAL}[LAL]{LIGO Algorithm Library}
\acrodef{GWTC}[GWTC]{Gravitational Wave Transient Catalog}

\newcommand{\PN}[0]{\ac{PN}\xspace}
\newcommand{\BBH}[0]{\ac{BBH}\xspace}
\newcommand{\BNS}[0]{\ac{BNS}\xspace}
\newcommand{\BH}[0]{\ac{BH}\xspace}
\newcommand{\NR}[0]{\ac{NR}\xspace}
\newcommand{\GW}[0]{\ac{GW}\xspace}
\newcommand{\SNR}[0]{\ac{SNR}\xspace}
\newcommand{\SSM}[0]{\ac{SSM}\xspace}
\newcommand{\aLIGO}[0]{\ac{aLIGO}\xspace}
\newcommand{\PSD}[0]{\ac{PSD}\xspace}
\newcommand{\GR}[0]{\ac{GR}\xspace}
\newcommand{\EOS}[0]{\ac{EOS}\xspace}
\newcommand{\LVC}[0]{\ac{LVC}\xspace}

% ======================
%  STYLIZED NAMES
% ======================

% Stylized pipeline names
\newcommand{\GSTLAL}{GstLAL\xspace}
\newcommand{\IMRPHENOMD}{IMRPhenomD\xspace}
\newcommand{\MANIFOLD}{{\fontfamily{qcr}\selectfont manifold}\xspace}
\newcommand{\SBANK}{{\fontfamily{qcr}\selectfont SBank}\xspace}

%% Capitalization
\newcommand\hmm[1]{\ifnum\ifhmode\spacefactor\else2000\fi>1500 \uppercase{#1}\else#1\fi}

\newcommand{\IGWNALERT}{\texttt{igwn-alert}\xspace}

%% MDC
\newcommand{\MDCSTART}{5 Jan. 2020 15:59:42}
\newcommand{\MDCEND}{14 Feb. 2020 15:59:42}

%% GSTLAL 
\newcommand{\TOTALTEMPLATES}{\ensuremath{1.8 \times 10^6}}
\newcommand{\CHECKERBOARDTEMPLATES}{\ensuremath{9 \times 10^5}}
\newcommand{\NUMSVDBANKS}{\ensuremath{\sim 1000}}
\newcommand{\TEMPLATESPERSUBBANK}{\ensuremath{\sim 500}}
\newcommand{\NUMSUBBANKSPERSVD}{\ensuremath{2}}
\newcommand{\SVDTOLERANCE}{\ensuremath{99.999\%}}
\newcommand{\PSDFFTLENGTH}{\ensuremath{4~\mathrm{seconds}}}
\newcommand{\FRAMELENGTH}{\ensuremath{1~\mathrm{second}}}
\newcommand{\BUFFERBLOCKSIZE}{\ensuremath{4096~\mathrm{bytes}}}
\newcommand{\FIRSTRIDE}{\ensuremath{0.25~\mathrm{seconds}}}
\newcommand{\TRIGGERSNRTHRESHOLD}{\ensuremath{4.0}}
\newcommand{\COINCTHRESHOLD}{\ensuremath{0.005~\mathrm{seconds}}}
\newcommand{\HTGATETHRESHOLDMIN}{\ensuremath{15.0}}
\newcommand{\HTGATETHRESHOLDMAX}{\ensuremath{100.0}}
\newcommand{\HTGATEMCHIRPMIN}{\ensuremath{0.8}}
\newcommand{\HTGATEMCHIRPMAX}{\ensuremath{45.0}}
\newcommand{\HTGATEMIN}{\ensuremath{\sim 15}}
\newcommand{\HTGATEMAX}{\ensuremath{\sim 325}}
\newcommand{\LRSNAPSHOT}{\ensuremath{4}}
\newcommand{\LRCOMPRESSION}{\ensuremath{0.003}}
\newcommand{\FARTRIALSFACTOR}{\ensuremath{2}}
\newcommand{\UPLOADCADENCE}{\ensuremath{4}}
\newcommand{\UPLOADCADENCEMDCTWELVE}{\ensuremath{2}}
\newcommand{\UPLOADDT}{\ensuremath{0.2}}
\newcommand{\SINGLESPENALTYMDCELEVEN}{\ensuremath{12}}
\newcommand{\SINGLESPENALTYOFOUR}{\ensuremath{13}}
\newcommand{\XISQMISMATCHRANGE}{\ensuremath{0.1-10\%}}

%% TEST SUITE
\newcommand{\TESTSUITECOINCWINDOW}{\ensuremath{\pm 1}}
\newcommand{\INJSNRFLOW}{\ensuremath{10.0}}
\newcommand{\INJSNRFHI}{\ensuremath{1600.0}}

%% SYMBOLS
\newcommand{\VT}{\ensuremath{\langle VT \rangle}}
\newcommand{\SPINZ}{\ensuremath{s_{i,z}}}
\newcommand{\CHIP}{\ensuremath{\chi_p}}
\newcommand{\MCHIRP}{\ensuremath{\mathcal{M}_c}\xspace}
\newcommand{\CHIEFF}{\ensuremath{\chi_{\mathrm{eff}}}}
\newcommand{\PASTRO}{\ensuremath{p(\mathrm{astro})}}
\newcommand{\MSUN}{\ensuremath{M_{\odot}}}
\newcommand{\TEND}{\ensuremath{t_{\mathrm{end}}}}

%% INJECTION SET
\newcommand{\TOTALINJECTIONS}{$5\times10^4$}
\newcommand{\BNSMAXZ}{\ensuremath{0.15}}
\newcommand{\NSBHMAXZ}{\ensuremath{0.25}}
\newcommand{\BBHMAXZ}{\ensuremath{1.9}}
\newcommand{\MDCDURATION}{\ensuremath{3.456\times10^6}}
\newcommand{\INJECTIONSPACING}{\ensuremath{\sim40}}

%% GWTC-3
\newcommand{\PASTROTHRESHOLD}{\ensuremath{0.50}}

%% INJECTION RECOVERY
\newcommand{\DECISIVESNRTHRESH}{\ensuremath{8.0}}
\newcommand{\NETWORKSNRTHRESH}{\ensuremath{10.0}}

\newcommand{\ALLABOVEDECSNRTHRESH}{\ensuremath{1522}}
\newcommand{\ALLINBANKABOVEDECSNRTHRESH}{\ensuremath{1457}}
\newcommand{\BBHINBANKABOVEDECSNRTHRESH}{\ensuremath{597}}
\newcommand{\BNSINBANKABOVEDECSNRTHRESH}{\ensuremath{482}}
\newcommand{\NSBHINBANKABOVEDECSNRTHRESH}{\ensuremath{378}}

\newcommand{\HIGHFARTHRESH}{$1$ per hour}
\newcommand{\LOWFARTHRESH}{$2$ per day}
\newcommand{\ONEPERHOUR}{\ensuremath{2.78\times10^{-4}~\mathrm{Hz}}}
\newcommand{\TWOPERDAY}{\ensuremath{2.31\times10^{-5}~\mathrm{Hz}}}
\newcommand{\ONEPERMONTH}{\ensuremath{3.85\times10^{-7}~\mathrm{Hz}}}
\newcommand{\TWOPERYEAR}{\ensuremath{3.16\times10^{-8}~\mathrm{Hz}}}

\newcommand{\ALLINBANKEFFICIENCY}[1]{%
	\IfEqCase{#1}{%
		{ONEPERHOUR}{\ensuremath{0.87}}%
		{TWOPERDAY}{\ensuremath{0.84}}%
		{ONEPERMONTH}{\ensuremath{0.78}}%
		{TWOPERYEAR}{\ensuremath{0.74}}%
	}[\PackageError{ALLINBANKEFFICIENCY}{Undefined option: #1}{}]
}%

\newcommand{\BBHINBANKEFFICIENCY}[1]{%
	\IfEqCase{#1}{%
		{ONEPERHOUR}{\ensuremath{0.87}}%
		{TWOPERDAY}{\ensuremath{0.84}}%
		{ONEPERMONTH}{\ensuremath{0.77}}%
		{TWOPERYEAR}{\ensuremath{0.71}}%
	}[\PackageError{BBHINBANKEFFICIENCY}{Undefined option: #1}{}]
}%

\newcommand{\BNSINBANKEFFICIENCY}[1]{%
	\IfEqCase{#1}{%
		{ONEPERHOUR}{\ensuremath{0.95}}%
		{TWOPERDAY}{\ensuremath{0.95}}%
		{ONEPERMONTH}{\ensuremath{0.89}}%
		{TWOPERYEAR}{\ensuremath{0.86}}%
	}[\PackageError{BNSINBANKEFFICIENCY}{Undefined option: #1}{}]
}%

\newcommand{\NSBHINBANKEFFICIENCY}[1]{%
	\IfEqCase{#1}{%
		{ONEPERHOUR}{\ensuremath{0.77}}%
		{TWOPERDAY}{\ensuremath{0.71}}%
		{ONEPERMONTH}{\ensuremath{0.65}}%
		{TWOPERYEAR}{\ensuremath{0.62}}%
	}[\PackageError{NSBHINBANKEFFICIENCY}{Undefined option: #1}{}]
}%

%% MDC12 INJECTION RECOVERY
\newcommand{\ALLABOVEDECSNRTHRESHMDCTWELVE}{\ensuremath{653}}
\newcommand{\ALLINBANKABOVEDECSNRTHRESHMDCTWELVE}{\ensuremath{621}}
\newcommand{\BBHINBANKABOVEDECSNRTHRESHMDCTWELVE}{\ensuremath{243}}
\newcommand{\BNSINBANKABOVEDECSNRTHRESHMDCTWELVE}{\ensuremath{209}}
\newcommand{\NSBHINBANKABOVEDECSNRTHRESHMDCTWELVE}{\ensuremath{169}}

\newcommand{\ALLINBANKEFFICIENCYMDCTWELVE}[1]{%
	\IfEqCase{#1}{%
		{ONEPERHOUR}{\ensuremath{0.88}}%
		{TWOPERDAY}{\ensuremath{0.86}}%
		{ONEPERMONTH}{\ensuremath{0.83}}%
		{TWOPERYEAR}{\ensuremath{0.81}}%
	}[\PackageError{ALLINBANKEFFICIENCYMDCTWELVE}{Undefined option: #1}{}]
}%

\newcommand{\BBHINBANKEFFICIENCYMDCTWELVE}[1]{%
	\IfEqCase{#1}{%
		{ONEPERHOUR}{\ensuremath{0.92}}%
		{TWOPERDAY}{\ensuremath{0.90}}%
		{ONEPERMONTH}{\ensuremath{0.87}}%
		{TWOPERYEAR}{\ensuremath{0.86}}%
	}[\PackageError{BBHINBANKEFFICIENCYMDCTWELVE}{Undefined option: #1}{}]
}%

\newcommand{\BNSINBANKEFFICIENCYMDCTWELVE}[1]{%
	\IfEqCase{#1}{%
		{ONEPERHOUR}{\ensuremath{0.95}}%
		{TWOPERDAY}{\ensuremath{0.93}}%
		{ONEPERMONTH}{\ensuremath{0.91}}%
		{TWOPERYEAR}{\ensuremath{0.88}}%
	}[\PackageError{BNSINBANKEFFICIENCYMDCTWELVE}{Undefined option: #1}{}]
}%

\newcommand{\NSBHINBANKEFFICIENCYMDCTWELVE}[1]{%
	\IfEqCase{#1}{%
		{ONEPERHOUR}{\ensuremath{0.74}}%
		{TWOPERDAY}{\ensuremath{0.72}}%
		{ONEPERMONTH}{\ensuremath{0.68}}%
		{TWOPERYEAR}{\ensuremath{0.66}}%
	}[\PackageError{NSBHINBANKEFFICIENCYMDCTWELVE}{Undefined option: #1}{}]
}%

%% PARAMETER ACCURACY
\newcommand{\MEAN}[1]{%
	\IfEqCase{#1}{%
		{MASSRATIO}{\ensuremath{1.39}}%
		{MCHIRP}{\ensuremath{0.15}}%
		{SPIN1Z}{\ensuremath{7.27}}%
		{SPIN2Z}{\ensuremath{2.81}}%
		{CHIEFF}{\ensuremath{5.77}}%
		{ENDTIME}{\ensuremath{6.23}}%
	}[\PackageError{MEAN}{Undefined option: #1}{}]
}%

\newcommand{\STDEV}[1]{%
	\IfEqCase{#1}{%
		{MASSRATIO}{\ensuremath{2.86}}%
		{MCHIRP}{\ensuremath{0.45}}%
		{SPIN1Z}{\ensuremath{285}}%
		{SPIN2Z}{\ensuremath{183}}%
		{CHIEFF}{\ensuremath{252}}%
		{ENDTIME}{\ensuremath{30.22}}%
	}[\PackageError{STDEV}{Undefined option: #1}{}]
}%

\newcommand{\BNSMCHIRPMEAN}{\ensuremath{2.06\times10^{-4}}}
\newcommand{\BNSMCHIRPSTDEV}{\ensuremath{8.33\times10^{-4}}}

\newcommand{\NSBHMCHIRPMEAN}{\ensuremath{-2.14\times10^{-4}}}
\newcommand{\NSBHMCHIRPSTDEV}{\ensuremath{6.26\times10^{-3}}}

\newcommand{\BBHMCHIRPMEAN}{\ensuremath{1.54\times10^{-1}}}
\newcommand{\BBHMCHIRPSTDEV}{\ensuremath{4.53\times10^{-1}}}

\newcommand{\BNSENDTIMEMEAN}{\ensuremath{-0.90}}
\newcommand{\BNSENDTIMESTDEV}{\ensuremath{18.0}}

\newcommand{\NSBHENDTIMEMEAN}{\ensuremath{18.7}}
\newcommand{\NSBHENDTIMESTDEV}{\ensuremath{59.3}}

\newcommand{\BBHENDTIMEMEAN}{\ensuremath{6.03}}
\newcommand{\BBHENDTIMESTDEV}{\ensuremath{11.3}}

\newcommand{\QFIFTY}[1]{%
	\IfEqCase{#1}{%
		{MASSRATIO}{\ensuremath{0.45}}%
		{MCHIRP}{\ensuremath{0.007}}%
		{SPIN1Z}{\ensuremath{1.24}}%
		{SPIN2Z}{\ensuremath{1.72}}%
		{CHIEFF}{\ensuremath{1.34}}%
		{ENDTIME}{\ensuremath{3.8}}%
	}[\PackageError{QFIFTY}{Undefined option: #1}{}]
}%

\newcommand{\QSEVENTYFIVE}[1]{%
	\IfEqCase{#1}{%
		{MASSRATIO}{\ensuremath{1.67}}%
		{MCHIRP}{\ensuremath{0.33}}%
		{SPIN1Z}{\ensuremath{3.82}}%
		{SPIN2Z}{\ensuremath{5.33}}%
		{CHIEFF}{\ensuremath{3.71}}%
		{ENDTIME}{\ensuremath{9.78}}%
	}[\PackageError{QSEVENTYFIVE}{Undefined option: #1}{}]
}%

\newcommand{\QNINETY}[1]{%
	\IfEqCase{#1}{%
		{MASSRATIO}{\ensuremath{4.97}}%
		{MCHIRP}{\ensuremath{0.73}}%
		{SPIN1Z}{\ensuremath{13.8}}%
		{SPIN2Z}{\ensuremath{17.5}}%
		{CHIEFF}{\ensuremath{10.8}}%
		{ENDTIME}{\ensuremath{25.6}}%
	}[\PackageError{QNINETY}{Undefined option: #1}{}]
}%

%% SEARCH SENSITIVITY
\newcommand{\GPCYRS}{\ensuremath{\mathrm{Gpc}^3\mathrm{yrs}}}
\newcommand{\INJECTEDVT}[1]{%
	\IfEqCase{#1}{%
		{BNS}{\ensuremath{1.08\times10^{-1}}}%
		{NSBH}{\ensuremath{4.34\times10^{-1}}}%
		{BBH}{\ensuremath{29.1}}%
	}[\PackageError{INJECTEDVT}{Undefined option: #1}{}]
}%

\newcommand{\VTTWOPERDAY}[1]{%
	\IfEqCase{#1}{%
		{BNS}{\ensuremath{3.49\times10^{-4}}}%
		{NSBH}{\ensuremath{8.08\times10^{-4}}}%
		{BBH}{\ensuremath{1.23\times10^{-1}}}%
	}[\PackageError{VTTWOPERDAY}{Undefined option: #1}{}]
}%

\newcommand{\VTNETSNR}[1]{%
	\IfEqCase{#1}{%
		{BNS}{\ensuremath{4.41\times10^{-4}}}%
		{NSBH}{\ensuremath{1.59\times10^{-3}}}%
		{BBH}{\ensuremath{1.52\times10^{-1}}}%
	}[\PackageError{VTNETSNR}{Undefined option: #1}{}]
}%

\newcommand{\VTDECSNR}[1]{%
	\IfEqCase{#1}{%
		{BNS}{\ensuremath{1.47\times10^{-4}}}%
		{NSBH}{\ensuremath{4.98\times10^{-3}}}%
		{BBH}{\ensuremath{5.46\times10^{-2}}}%
	}[\PackageError{VTDECSNR}{Undefined option: #1}{}]
}%

%% SKY LOCALIZATION

\newcommand{\SEARCHEDAREAQFIFTY}[1]{%
	\IfEqCase{#1}{%
		{ALL}{\ensuremath{271}}%
		{TRIPLE}{\ensuremath{31.9}}%
		{DOUBLE}{\ensuremath{301}}%
		{SINGLE}{\ensuremath{3150}}%
	}[\PackageError{SEARCHEDAREAQFIFTY}{Undefined option: #1}{}]
}%

\newcommand{\SEARCHEDAREAQSEVENTYFIVE}[1]{%
	\IfEqCase{#1}{%
		{ALL}{\ensuremath{1080}}%
		{TRIPLE}{\ensuremath{140}}%
		{DOUBLE}{\ensuremath{893}}%
		{SINGLE}{\ensuremath{10,400}}%
	}[\PackageError{SEARCHEDAREAQSEVENTYFIVE}{Undefined option: #1}{}]
}%

\newcommand{\SEARCHEDAREAQNINETY}[1]{%
	\IfEqCase{#1}{%
		{ALL}{\ensuremath{3910}}%
		{TRIPLE}{\ensuremath{357}}%
		{DOUBLE}{\ensuremath{2470}}%
		{SINGLE}{\ensuremath{18,400}}%
	}[\PackageError{SEARCHEDAREAQNINETY}{Undefined option: #1}{}]
}%

\newcommand{\SEARCHEDPROBQFIFTY}[1]{%
	\IfEqCase{#1}{%
		{ALL}{\ensuremath{0.53}}%
		{TRIPLE}{\ensuremath{0.58}}%
		{DOUBLE}{\ensuremath{0.52}}%
		{SINGLE}{\ensuremath{0.59}}%
	}[\PackageError{SEARCHEDPROBQFIFTY}{Undefined option: #1}{}]
}%

\newcommand{\SEARCHEDPROBQSEVENTYFIVE}[1]{%
	\IfEqCase{#1}{%
		{ALL}{\ensuremath{0.79}}%
		{TRIPLE}{\ensuremath{0.84}}%
		{DOUBLE}{\ensuremath{0.77}}%
		{SINGLE}{\ensuremath{0.78}}%
	}[\PackageError{SEARCHEDPROBQSEVENTYFIVE}{Undefined option: #1}{}]
}%

\newcommand{\SEARCHEDPROBQNINETY}[1]{%
	\IfEqCase{#1}{%
		{ALL}{\ensuremath{0.93}}%
		{TRIPLE}{\ensuremath{0.96}}%
		{DOUBLE}{\ensuremath{0.92}}%
		{SINGLE}{\ensuremath{0.92}}%
	}[\PackageError{SEARCHEDPROBQNINETY}{Undefined option: #1}{}]
}%

%% PASTRO 
\newcommand{\BNSTOBNS}{\ensuremath{90.3\%}}
\newcommand{\BNSTONSBH}{\ensuremath{9.7\%}}

\newcommand{\NSBHTONSBH}{\ensuremath{64.1\%}}
\newcommand{\NSBHTOBBH}{\ensuremath{33.8\%}}
\newcommand{\NSBHTOBNS}{\ensuremath{2.10\%}}

\newcommand{\BBHTOBBH}{\ensuremath{100\%}}

\newcommand{\TERRTOTERR}{\ensuremath{2.60\%}}
\newcommand{\TERRTOBBH}{\ensuremath{68.8\%}}
\newcommand{\TERRTONSBH}{\ensuremath{19.5\%}}
\newcommand{\TERRTOBNS}{\ensuremath{9.10\%}}

%% MDC12 PASTRO 
\newcommand{\BNSTOBNSMDCTWELVE}{\ensuremath{79.8\%}}
\newcommand{\BNSTONSBHMDCTWELVE}{\ensuremath{20.2\%}}

\newcommand{\NSBHTONSBHMDCTWELVE}{\ensuremath{92.1\%}}
\newcommand{\NSBHTOBBHMDCTWELVE}{\ensuremath{6.83\%}}
\newcommand{\NSBHTOBNSMDCTWELVE}{\ensuremath{1.02\%}}

\newcommand{\BBHTOBBHMDCTWELVE}{\ensuremath{99.5\%}}
\newcommand{\BBHTONSBHMDCTWELVE}{\ensuremath{0.05\%}}

\newcommand{\TERRTOBBHMDCTWELVE}{\ensuremath{76.2\%}}
\newcommand{\TERRTONSBHMDCTWELVE}{\ensuremath{23.8\%}}

%% GRAVITATIONAL WAVE RECOVERY
\newcommand{\OTHREEOPA}{\ensuremath{1.2}} % FAR (per year)

\newcommand{\MDCGWIFOS}[1]{%
	\IfEqCase{#1}{%
		{GW200112}{L1}%
		{GW200115}{H1L1}%
		{GW200128}{H1L1}%
		{GW200129}{H1L1V1}%
		{GW200202}{H1L1}%
		{GW200208q}{H1L1}%
		{GW200208am}{H1L1}%
		{GW200209}{H1L1}%
		{GW200210}{H1L1}%
	}[\PackageError{MDCGWIFOS}{Undefined option: #1}{}]
}%

\newcommand{\MDCGWSNR}[1]{%
	\IfEqCase{#1}{%
		{GW200112}{\ensuremath{18.46}}%
		{GW200115}{\ensuremath{11.48}}%
		{GW200128}{\ensuremath{9.98}}%
		{GW200129}{\ensuremath{26.30}}%
		{GW200202}{\ensuremath{11.09}}%
		{GW200208q}{\ensuremath{10.56}}%
		{GW200208am}{\ensuremath{8.00}}%
		{GW200209}{\ensuremath{9.96}}%
		{GW200210}{\ensuremath{9.28}}%
	}[\PackageError{MDCGWSNR}{Undefined option: #1}{}]
}%

% FAR (yr-1)
\newcommand{\MDCGWFAR}[1]{%
	\IfEqCase{#1}{%
		{GW200112}{\ensuremath{1.01\times10^{-7}}}%
		{GW200115}{\ensuremath{2.55\times10^{-4}}}%
		{GW200128}{\ensuremath{1.44\times10^{-4}}}%
		{GW200129}{\ensuremath{1.78\times10^{-17}}}%
		{GW200202}{\ensuremath{1.69\times10^{-2}}}%
		{GW200208q}{\ensuremath{4.92\times10^{-5}}}%
		{GW200208am}{\ensuremath{2.02\times10^{3}}}%
		{GW200209}{\ensuremath{1.20}}%
		{GW200210}{\ensuremath{3.64\times10^{3}}}%
	}[\PackageError{MDCGWFAR}{Undefined option: #1}{}]
}%

\newcommand{\MDCGWPASTRO}[1]{%
	\IfEqCase{#1}{%
		{GW200112}{\ensuremath{>0.99}}%
		{GW200115}{\ensuremath{>0.99}}%
		{GW200128}{\ensuremath{>0.99}}%
		{GW200129}{\ensuremath{>0.99}}%
		{GW200202}{\ensuremath{>0.99}}%
		{GW200208q}{\ensuremath{>0.99}}%
		{GW200208am}{\ensuremath{0.48}}%
		{GW200209}{\ensuremath{>0.99}}%
		{GW200210}{0.27}%
	}[\PackageError{MDCGWPASTRO}{Undefined option: #1}{}]
}%

\newcommand{\MDCGWMCHIRP}[1]{%
	\IfEqCase{#1}{%
		{GW200112}{\ensuremath{33.37~M_{\odot}}}%
		{GW200115}{\ensuremath{2.58~M_{\odot}}}%
		{GW200128}{\ensuremath{50.74~M_{\odot}}}%
		{GW200129}{\ensuremath{30.66~M_{\odot}}}%
		{GW200202}{\ensuremath{8.15~M_{\odot}}}%
		{GW200208q}{\ensuremath{34.50~M_{\odot}}}%
		{GW200208am}{\ensuremath{66.59~M_{\odot}}}%
		{GW200209}{\ensuremath{39.45~M_{\odot}}}%
		{GW200210}{\ensuremath{7.89~M_{\odot}}}%
	}[\PackageError{MDCGWMCHIRP}{Undefined option: #1}{}]
}%

\newcommand{\OTHREEGWIFOS}[1]{%
	\IfEqCase{#1}{%
		{GW200112}{L1}%
		{GW200115}{H1L1}%
		{GW200128}{--}%
		{GW200129}{H1L1V1}%
		{GW200202}{--}%
		{GW200208q}{--}%
		{GW200208am}{--}%
		{GW200209}{--}%
		{GW200210}{--}%
	}[\PackageError{OTHREEGWIFOS}{Undefined option: #1}{}]
}%

\newcommand{\OTHREEGWSNR}[1]{%
	\IfEqCase{#1}{%
		{GW200112}{\ensuremath{18.79}}%
		{GW200115}{\ensuremath{11.42}}%
		{GW200128}{--}%
		{GW200129}{\ensuremath{26.61}}%
		{GW200202}{--}%
		{GW200208q}{--}%
		{GW200208am}{--}%
		{GW200209}{--}%
		{GW200210}{--}%
	}[\PackageError{OTHREEGWSNR}{Undefined option: #1}{}]
}%

% FAR (yr-1)
\newcommand{\OTHREEGWFAR}[1]{%
	\IfEqCase{#1}{%
		{GW200112}{\ensuremath{4.05\times10^{-4}}}%
		{GW200115}{\ensuremath{6.61\times10^{-4}}}%
		{GW200128}{\ensuremath{> \OTHREEOPA{}}}%
		{GW200129}{\ensuremath{2.11\times10^{-24}}}%
		{GW200202}{\ensuremath{> \OTHREEOPA{}}}%
		{GW200208q}{--}%
		{GW200208am}{--}%
		{GW200209}{--}%
		{GW200210}{--}%
	}[\PackageError{OTHREEGWFAR}{Undefined option: #1}{}]
}%

\newcommand{\OTHREEGWPASTRO}[1]{%
	\IfEqCase{#1}{%
		{GW200112}{\ensuremath{>0.99}}%
		{GW200115}{\ensuremath{>0.99}}%
		{GW200128}{--}%
		{GW200129}{\ensuremath{>0.99}}%
		{GW200202}{--}%
		{GW200208q}{--}%
		{GW200208am}{--}%
		{GW200209}{--}%
		{GW200210}{--}%
	}[\PackageError{OTHREEGWPASTRO}{Undefined option: #1}{}]
}%

\newcommand{\OTHREEGWMCHIRP}[1]{%
	\IfEqCase{#1}{%
		{GW200112}{\ensuremath{35.37~M_{\odot}}}%
		{GW200115}{\ensuremath{2.57~M_{\odot}}}%
		{GW200128}{--}%
		{GW200129}{\ensuremath{32.74~M_{\odot}}}%
		{GW200202}{--}%
		{GW200208q}{--}%
		{GW200208am}{--}%
		{GW200209}{--}%
		{GW200210}{--}%
	}[\PackageError{OTHREEGWMCHIRP}{Undefined option: #1}{}]
}

%% RETRACTIONS
\newcommand{\OTHREERETRACTIONS}{23}
\newcommand{\OTHREEGSTLALRETRACTIONS}{15}

\newcommand{\RETRACTIONFAR}{\ensuremath{1.67~\mathrm{per}~\mathrm{year}}}
\newcommand{\RETRACTIONSNR}{\ensuremath{14.5}}
\newcommand{\MDCRETRACTIONFARTHRESH}{one per year}

%% TEMPLATE BANK
\newcommand{\BANKMASSLOW}{\ensuremath{1.0~M_{\odot}}}
\newcommand{\BANKMASSHIGH}{\ensuremath{200~M_{\odot}}}

\newcommand{\BHMASSLOW}{\ensuremath{3.0~M_{\odot}}}
\newcommand{\NSMASSLOW}{\ensuremath{1.0~M_{\odot}}}
\newcommand{\NSMASSHIGH}{\ensuremath{3.0 M_{\odot}}}
\newcommand{\TOTALMASSHIGH}{\ensuremath{400.0~M_{\odot}}}
\newcommand{\MASSRATIOHIGH}{\ensuremath{20}}

\newcommand{\NSSPIN}{\ensuremath{0.05}}
\newcommand{\BHSPIN}{\ensuremath{0.99}}
\newcommand{\CHIPBOUND}{\ensuremath{1\times10^{-3}}}

\newcommand{\MCHIRPBOUNDARY}{\ensuremath{1.73~M_{\odot}}}
\newcommand{\LOWMCHIRPWAVEFORM}{\texttt{TaylorF2}}
\newcommand{\HIGHMCHIRPWAVEFORM}{\texttt{SEOBNRv4}}

%% LATENCY
\newcommand{\MDCELEVENLOFARLATENCY}{\ensuremath{14.58}}
\newcommand{\MDCELEVENHIFARLATENCY}{\ensuremath{10.30}}

\newcommand{\MDCTWELVELOFARLATENCY}{\ensuremath{12.04}}
\newcommand{\MDCTWELVEHIFARLATENCY}{\ensuremath{9.86}}

\section{Introduction}
\label{sec:introduction}
Since the \ac{O2} of \ac{LVK}, \acp{GW} have emerged as an important messenger
in multi-messenger astronomy. It was during this observing run that GW170817 was
detected~\cite{GCN21505,LIGOScientific:2017vwq,LIGOScientific:2018mvr}. The source 
of this event was a \ac{BNS}, and thus electromagnetically bright, leading to the
first multi-messenger detection involving \acp{GW}. It has since led to a
variety of new scientific results~\cite{PhysRevLett.121.161101, PhysRevLett.123.011102}.

Participation of \acp{GW} in multi-messenger detections like GW170817 is made possible
by a combination of multiple analysis pipelines and tools working together. First, a \ac{GW}
search pipeline analyzes the strain data produced by \ac{GW} detectors like the \ac{LIGO}
Hanford and Livingston detectors~\cite{ligo}, the Virgo detector~\cite{virgo},
and the KAGRA detector~\cite{kagra}, finds \ac{GW} candidates in the data, and uploads them
to the Gravitational Wave Candidate Event Database (GraceDB)~\cite{gracedb} in near-real
time. Examples of \ac{GW} search pipelines are
GstLAL~\cite{Messick:2016aqy, Cannon:2020qnf, Sachdev:2019vvd, Hanna:2019ezx},
IAS~\cite{ias, Zackay_2021}, MBTA~\cite{mbta, Adams_2016},
PyCBC~\cite{Dal_Canton_2021, Davies_2020, pycbc}, and SPIIR~\cite{spiir, spiir_2017}.

\begin{figure*}
\includegraphics[width=\textwidth]{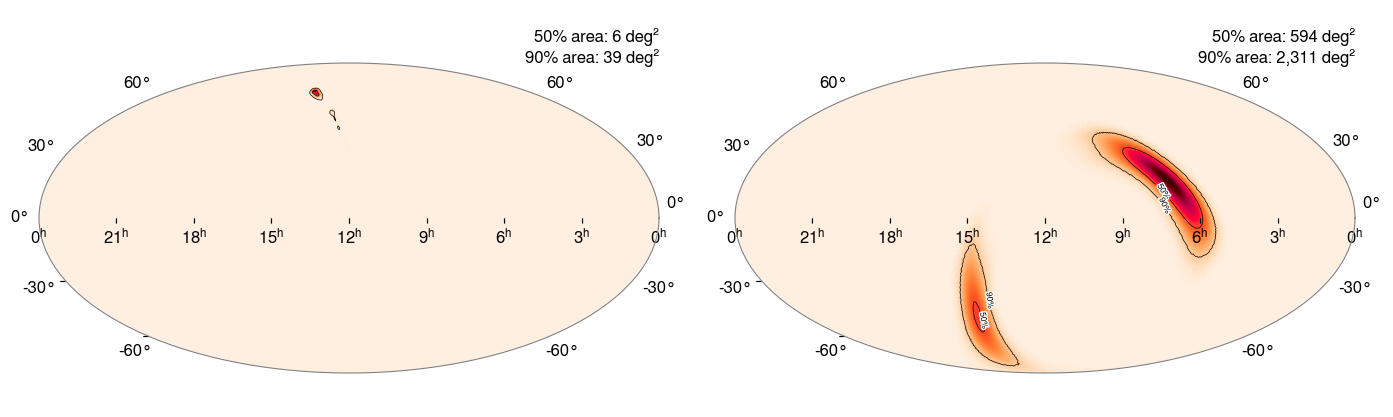}
\caption{\label{fig:skymap}
This figure shows two examples of sky maps. The one on the left is relatively
well constrained in terms of sky location, while the one of the right is
less constrained. Louder \ac{GW} signals being detected in more number
of detectors produces more constrained, and hence better sky maps. These sky
maps were produced using the BAYESTAR package on simulated \ac{GW} signals.
}
\end{figure*}

Next, \ac{GW} inference pipelines ingest the results of \ac{GW} searches, and
infer source properties like the source parameters, sky location, distance, etc.
Sky location is generally communicated in the form of a sky map, a two-dimensional
plot of the sky showing contours for probable location of the source. Two commonly
used contour values are 50\% and 90\%. Examples of sky maps are shown in 
\figref{fig:skymap}. LALInference~\cite{LSCAlgorithmLibrary,PhysRevD.88.062001},
BAYESTAR~\cite{Singer:2015ema, Singer:2016eax}, and
BILBY~\cite{Ashton:2018jfp, Romero-Shaw:2020owr} are some of the inference
pipelines used.

After the sky map for a \ac{GW} candidate is calculated, a public alert gets sent~\cite{gcn}.
All of this happens within seconds to minutes of the \ac{GW} signal reaching Earth.
Astronomers can then choose to point their telescopes at the location described by the
sky map included in the public alert, to try to observe any electromagnetic counterparts
to the \ac{GW} candidate. However, these electromagnetic counterparts can be very faint
and can fade within seconds of the \ac{GW} detection. Because of this, it is crucial for
multi-messenger efforts to extract all possible information from the \ac{GW} signal
to give astronomers the best opportunity to observe any electromagnetic counterparts.

How well constrained the source is in a sky map depends very heavily on the number of 
detectors contributing to the detection of the candidate. A single detector gives
almost no information. Two detectors generally localize the source to a partially filled circle in the sky.
Three detectors, however, can localize the source to a few small regions in the sky, in the best cases. For example,
in \figref{fig:skymap}, the left panel is derived from a candidate comprising three
detectors, whereas the right panel is derived from a candidate comprising two. With more than
three detectors, even smaller regions can be obtained. Consequently, efforts are constantly
taken to minimize detector downtime, and to coordinate maintenance schedules among
among detectors to maximize coincident observing time.

Low-latency inference pipelines like BAYESTAR extract sky location information
by looking at the arrival times, phases, and amplitudes of the \ac{GW} signal at the different detectors,
as well as their evolution in time around the time of the candidate,
as reported by the \ac{GW} search. Consequently, by measuring these values more
accurately, \ac{GW} searches can contribute to better sky maps. 

In this work, we introduce the GstLAL SNR Optimizer, which is designed to do just that.
GstLAL, like other modeled \ac{GW} searches uses a collection of \ac{GW} template waveforms, each with
a different combination of source parameters, called a ``template bank", and correlates each
one of them across the \ac{GW} strain data, in a process called matched filtering. The
output of this process is called the \ac{SNR} time series. However, since the parameter
space of \ac{GW} sources is only discretely sampled by the template bank, we expect
a loss in the \ac{SNR}, and in the accuracy of the arrival times and phases measured by
GstLAL. Common values of the minimal match between neighboring templates in a template
bank range from 0.97 to 0.99~\cite{Sakon:2022ibh}. As a result, we expect around a 1 to 3\%
loss in the SNR. The GstLAL SNR Optimizer, referred to as the SNR Optimizer hereafter, ingests
GstLAL low latency results in real time, and performs a small targeted hierarchical search.
The search is targeted in two ways: The SNR Optimizer only analyzes the data close in time
to a candidate reported by GstLAL, and it also analyzes the parameter space close to
the parameters reported by GstLAL. It does not analyze a fixed bank of templates, but
rather dynamically creates new ones in the relevant parameter space, hierarchically
closing in on the true location of the signal in the parameter space. It also implements
other improvements, like more accurate matched filtering, leading to higher \acp{SNR},
and hence better sky maps. It does all this in a timescale of a few seconds to five
minutes, and if it manages to produce better results, the skymap produced from its
results is included in the public alert issues for the candidate.

In \secref{sec:methodology}, we will discuss the detailed implementation of the 
SNR Optimizer, and in \secref{sec:results}, we will discuss the tests that we ran
to measure the performance of the SNR Optimizer.

\section{Methodology}
\label{sec:methodology}

\subsection{General GstLAL Methods}
As described before, modeled \ac{GW} searches like GstLAL make use of a template bank that
discretely samples the intrinsic parameter space (i.e. masses and spins) of the \ac{GW}
sources, causing a loss in \ac{SNR}. Before the data can be matched filtered with the templates, both the data
and the templates are whitened using a \ac{PSD} that represents the frequency characteristics
of detector noise. GstLAL implements whitening in the frequency domain, and matched filtering
in the time domain. The whitened data is defined as
\begin{equation}
\label{eq:data_whitening}
\hat{d}(\tau) = \int_{-\infty}^{\infty} df \frac{\tilde{d}(f)}{\sqrt{S_{n}(|f|)/2}} e^{2\pi if\tau}
\end{equation}
where $\tilde{d}(f)$ is the data represented in the frequency space, and $S_{n}(|f|)$ is the
single-sided \ac{PSD}. Similarly, the whitened template is defined as
\begin{equation}
\label{eq:template_whitening}
\hat{h}(\tau) = \int_{-\infty}^{\infty} df \frac{\tilde{h}(f)}{\sqrt{S_{n}(|f|)/2}} e^{2\pi if\tau}
\end{equation}
where $\tilde{h}(f)$ is the template represented in the frequency space.
The matched filter output for that particular template, i.e. the \ac{SNR} is then calculated as
\begin{equation}
\label{eq:snr}
\mathrm{SNR}(t) = \int_{-\infty}^{\infty} d\tau \hat{d}(t + \tau) \hat{h}(\tau)
\end{equation}

GstLAL currently has the ability to whiten the data in real time, but template whitening
needs to be done before starting the process of matched filtering. For a low-latency analysis,
since the data are not known beforehand, this means that the templates are whitened using
a PSD projected to represent future detector noise~\cite{joshi2025timesmatchedfiltergravitational}. This mismatch
between the PSD used to whiten the templates, and the ``true" PSD measured from detector noise
also causes a loss in \ac{SNR}.

Since ensuring the least possible latencies for public alerts is crucial for multi-messenger astronomy,
the GstLAL low-latency analysis makes certain concessions on the quality of matched filtering.
These include using a relatively short length of data (i.e. an FFT length of 4 seconds of data)
for measuring the \ac{PSD} and whitening the data in \eqref{eq:data_whitening}. This results in
lower statistics while measuring the \ac{PSD}, and produces a \ac{PSD} sampled at relatively larger
intervals of frequency. Similarly, the data itself is sampled at a relatively lower rate of 2048 Hz.
Both of these concessions case a decrease in the \ac{SNR}.

The GstLAL analysis recognizes times during which the \ac{SNR} of some template crosses the
threshold value of 4, called a ``trigger". Triggers recognized as originating from noise~\cite{Joshi:2023ltf} are
added to the background. Triggers are then ranked against this background, and a likelihood
ratio (LR)~\cite{Tsukada:2023} is calculated as a ranking statistic. Triggers with high LRs
are called \ac{GW} candidates. The LR is then converted
to a \ac{FAR} by taking into account the LR statistics of noise triggers, as well the livetime
of the analysis. For a low-latency analysis, if the \ac{FAR} crosses a specific threshold~\cite{opa},
the candidate is uploaded to GraceDB, and a public alert is issued~\cite{gcn}.
The public alert contains information about the sky location of the source, low-latency
parameter estimation of the source~\cite{rose2024rapid}, as well as information about
the probability of astrophysical origin for different source classes~\cite{Ray:2023nhx, villa2022astrophysical, Andres_2022}.
All of this can help astronomers follow-up on any potential electromagnetic counterparts.
A detailed description of the
GstLAL low-latency analysis can be found in~\cite{Ewing:2023}.

In summary, a GstLAL low-latency analysis loses \ac{SNR} due to the following reasons:
\begin{enumerate}
\item{discrete nature of the fixed template bank}
\item{templates not whitened with a PSD measured in real time}
\item{low FFT length for \ac{PSD} measurement and data whitening}
\item{low rate of data sampling}
\end{enumerate}

\subsection{SNR Optimizer Methods}
\subsubsection{Design Principles}
The SNR Optimizer is designed to mitigate any loss in SNR due to the aforementioned
reasons, and maximize SNR, leading to better sky maps. It implements a targeted,
hierarchical, and sub-threshold search in order to follow-up on candidates reported by GstLAL, and 
recover them with a higher SNR. The SNR Optimizer has evolved from the metric assisted
stochastic sampling (MASS) \ac{GW} search described in~\cite{PhysRevD.106.084033}.
As such, the \ac{SNR} calculation for any given template is identical to that performed
by MASS. A central philosophy of the SNR Optimizer
is that it does not attempt to evaluate the significance of a candidate. Instead it
relies on the GstLAL low-latency search not only to provide it with candidates that arise
from astrophysical sources rather than noise, but also to provide the LR, \ac{FAR}, and
the probability of astrophysical origin for different source classes for
these candidates.

\subsubsection{Template Bank}
The SNR Optimizer does not rely on a fixed bank of pre-created templates. Instead, because of
its hierarchical nature, it starts off with a fixed template bank, and dynamically creates 
new templates in between the gaps of the original templates, as it closes in on the true
location of the signal in the intrinsic parameter space. To do this, it makes use of template
banks created by the \texttt{manifold}~\cite{PhysRevD.108.042003} software package.

\begin{figure}
\includegraphics[width=0.5\columnwidth]{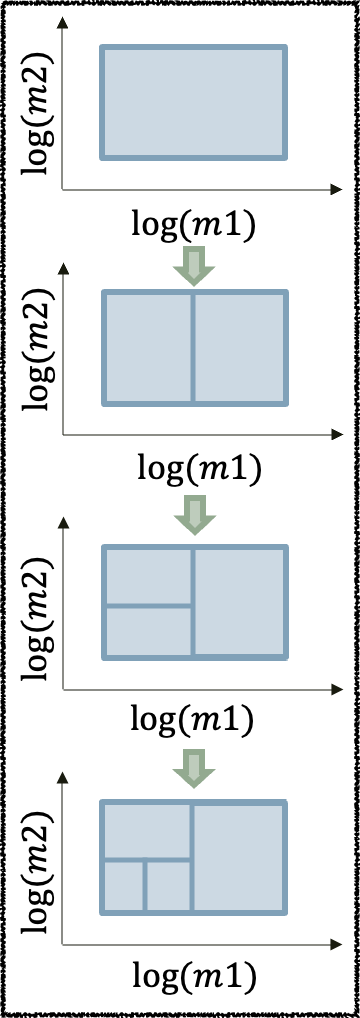}
\caption{\label{fig:split}
This figure shows a schematic of the template splitting process enabled by the manifold rectangle structure.
Each rectangle has a template at its center, and the metric is calculated at that location. The boundaries
of the rectangle represent the area (or volume, depending on the dimensionality of the space) occupied by
the template. This schematic shows a template splitting into two new templates, one of which again splits
into two new templates, and again one more time. This schematic shows templates only in the $m_1$-$m_2$ space.
For the fourth observing run, the SNR Optimizer operates in the $m_1$-$m_2$-$\chi_{eff}$ space.
}
\end{figure}

\texttt{Manifold} implements a metric on the intrinsic parameter space in order to place templates at a fixed
mismatch from each other, and this metric information at the location of every template is stored
in the template bank file. Templates are expressed as hyper-rectangles in the intrinsic parameter space.
Each such rectangle comprises a fixed area in the intrinsic parameter space (depending on the dimensionality of
the space, this might actually be a 3D or 4D volume), the template at the
center of this area, and the metric at the location of the template
Using the metric, any template can be split into two new templates, each occupying
half the area of the original template. The metric gets re-calculated at the
locations of the new templates enabling this process to be repeated indefinitely.
This operation is cheap, since the calculation is done only approximately,
enabling information from the metric at the original template to be re-used.
This process is illustrated in \figref{fig:split}.
This easy splitting of templates into multiple new ones enabled by \texttt{manifold}
facilitates the hierarchical nature of the SNR Optimizer.

\subsubsection{Search Algorithm}
Similar to GstLAL, the SNR Optimizer can be run in a low-latency ``online" mode, or a high-latency
``offline" mode. In the offline mode, it reads in a gstlal event file for a particular candidate,
and analyzes a small amount of data around it. The offline mode is meant only for testing purposes.
More importantly, in the online mode, the SNR Optimizer continuously reads in live data,
continuously calculates the \ac{PSD} from the data, and whitens it with that \ac{PSD}.
Because it operates in a medium-latency
timescale, it can afford to use a higher FFT length than GstLAL for the purposes of
data whitening. It can also sample the data at higher rate than GstLAL, since it does not matched filter
all of the data, but rather only short amounts of data around candidates, and hence there is no
danger of falling behind live data. Both of these factors contribute to the higher SNRs obtained
by the SNR optimizer.

Similar to GstLAL, the SNR Optimizer also implements ``gating" on the
whitened data, in which if the amplitude of whitened data exceeds a certain number of standard
deviations, it is instead set to 0, with 0.25 seconds of padding on either side~\cite{Messick:2016aqy, Sachdev:2019vvd}.
This is done to remove any transient non-gaussian components in the noise, called ``glitches",
which can artificially increase \ac{SNR} and mimic \acp{GW}.

While it is doing this, it also listens for any candidates that the GstLAL low-latency analysis
reports. When it receives such a candidate, it performs matched filtering on the data in a 2 second
window (plus sufficient padding on either side) around the event time as reported by GstLAL. Because
it was continuously whitening data before, data whitening does not need to be done from scratch for
the candidate, helping with latencies. It does not filter the data using all the templates in the bank,
but rather chooses the 1000 closest templates to the template reported by GstLAL. Because the metric
at each template it known, the operation for finding the closest templates to a given template is a cheap one.
The underlying assumption in only choosing 1000 templates to filter is in accordance with the central
philosophy of the SNR Optimizer, that GstLAL will only provide astrophysical candidates to the SNR Optimizer,
and as such the template reported by GstLAL is close to the ``true" template.
All templates are whitened in real time, using the latest estimate of the \ac{PSD}.
In this way, the SNR Optimizer sets up a targeted search around the candidate.

After it performs matched filtering for those 1000 templates, triggers are formed from all 1000 templates
using \acp{SNR} from all detectors. Out of those, only 500 templates with the highest network SNR
are retained, while the rest are discarded. These 500 templates are then each split into 2 new
templates, and the SNR at each of these new templates is calculated, and triggers are formed.
This cycle is repeated, now using the 500 retained templates, as well as their child templates, and 
this time around, only 250 templates get retained.

There are 2 stopping conditions to this loop. A template will not split further if its area is lower
than some threshold value. This would mean that further splitting it would only give templates very
close by, causing diminishing returns. If at any iteration of the loop, no new templates can be formed
by this process of splitting, that means we have arrived at our final set of templates, and the
template whose trigger has the highest network SNR is chosen as the template and the candidate
reported by the SNR Optimizer. The second stopping condition is if the SNR Optimizer reaches
4 minutes of processing time from the time it received the candidate from GstLAL. When that happens,
the SNR Optimizer takes whatever set of templates it had retained up to that point, and reports
the template with the highest network SNR trigger as its template and candidate. The reason for
this timeout is that the \ac{LVK} has an internal timeout of 4.5 minutes~\cite{gwcelery} for candidates
to be considered for generating a sky map and issuing a public alert. In this way, the SNR 
Optimizer implements a hierarchical search to recover the candidate with a template very close to
the true template of the \ac{GW} signal.

\subsubsection{Coincidence Formation}
The method implemented by the SNR Optimizer for forming coincidences across detectors when
creating a candidate is different than that implemented by GstLAL, because of two reasons:
\begin{enumerate}
\item{The SNR Optimizer is a sub-threshold search, whereas GstLAL is not. This means that the SNR Optimizer
can create triggers even below an SNR of 4, unlike GstLAL, and consider them for forming coincidences.}
\item{Since GstLAL calculates a LR based on the properties of the coincidence itself, it can afford
to be less strict when forming coincidences, because if it forms a ``bad" coincidence (i.e. with
an unphysical combination of arrival times, phases and \acp{SNR} at different detectors), the part
of the LR equation that calculates the probability of such a combination of times, phases and \acp{SNR},
commonly called the $dt-d\phi-dSNR$ term of the LR, will downrank this trigger, and it will not become
a candidate. However, since the SNR Optimizer does not do any significance estimation itself, it needs to
make sure the coincidence it forms is physical.}
\end{enumerate}

In order to create coincidences in a sub-threshold search, the SNR Optimizer first finds peaks in
the \ac{SNR} timeseries for every detector independently, in the 2 second window of data being analyzed.
It then loops over these peaks, and for every one, it adds the other detectors by finding new peaks 
in such a way that the combination of arrival times at the detectors remains physical. No arrival phase
or SNR information is used to assess the physicality of the coincidence, and in the next paragraph,
we will show that this is enough to guarantee physical coincidences. Whenever new detectors
are being added, the order of addition is kept the same as the descending order of SNRs of the original
\ac{SNR} peaks. This is done because we want to add louder detectors earlier, when there are less
constraints on them for keeping the coincidence physical. After the loop over all detectors is finished
we are left with coincidences equal in number to the number of detectors. The coincidence with
the maximum network \ac{SNR} is then chosen to be the candidate reported by the SNR Optimizer for that
window of data.

The two \ac{LIGO} detectors and the Virgo detector form a plane, and hence the arrival times of a
candidate at the detectors can be converted into information of the arrival direction of the \ac{GW}
and its velocity parallel to this plane, by solving the following system of equations:
\begin{align}
t_1 &= \frac{\mathbf{n} \cdot \mathbf{r_1}}{c} \\
t_2 &= \frac{\mathbf{n} \cdot \mathbf{r_2}}{c} \\
t_3 &= \frac{\mathbf{n} \cdot \mathbf{r_3}}{c} 
\end{align}
where $t_i$ are the arrival times at the detectors, $\mathbf{n}$ is the vector representing the direction of the \ac{GW}, and $r_i$ are the
location vectors of the three detectors.
If the velocity of the \ac{GW} parallel to the plane so obtained is less than or equal to c, the speed of light,
the combination of times are physical. This argument can be condensed into the calculation of a
$\chi^2$ statistic:
\begin{equation}
\chi^2 = (t_1 - \frac{\mathbf{n_{\parallel}} \cdot \mathbf{r_1}}{c} - t)^2 + (t_2 - \frac{\mathbf{n_{\parallel}} \cdot \mathbf{r_2}}{c} - t)^2 + (t_3 - \frac{\mathbf{n_{\parallel}} \cdot \mathbf{r_3}}{c} - t)^2
\end{equation}
where t is the average arrival time that minimizes the expression.
For a physical combination of times, $\chi^2$ will be 0, whereas
for an unphysical combination, it will be larger than 0. While forming coincidences, only
those arrival times that give a $\chi^2$ less than 3 are used. When adding a second detector to
a coincidence, this just means selecting the highest \ac{SNR} peak within the light travel time
of the two detectors. However, when adding the third detector, all possible sample points
within the light travel time of the two earlier detectors are considered, and the maximum
\ac{SNR} one that gives a $\chi^2$ less than 3 is added. The reason for selecting a $\chi^2$
threshold of 3 instead of 0 is that due to numerical noise, even a physical combination of 
time will not give a $\chi^2$ perfectly equal to 0, but rather a very small value. A second reason
is that due to detector noise, the arrival time estimate might not be perfect, and the $\chi^2$
threshold needs to allow for that.

\begin{figure}
\includegraphics[width=\columnwidth]{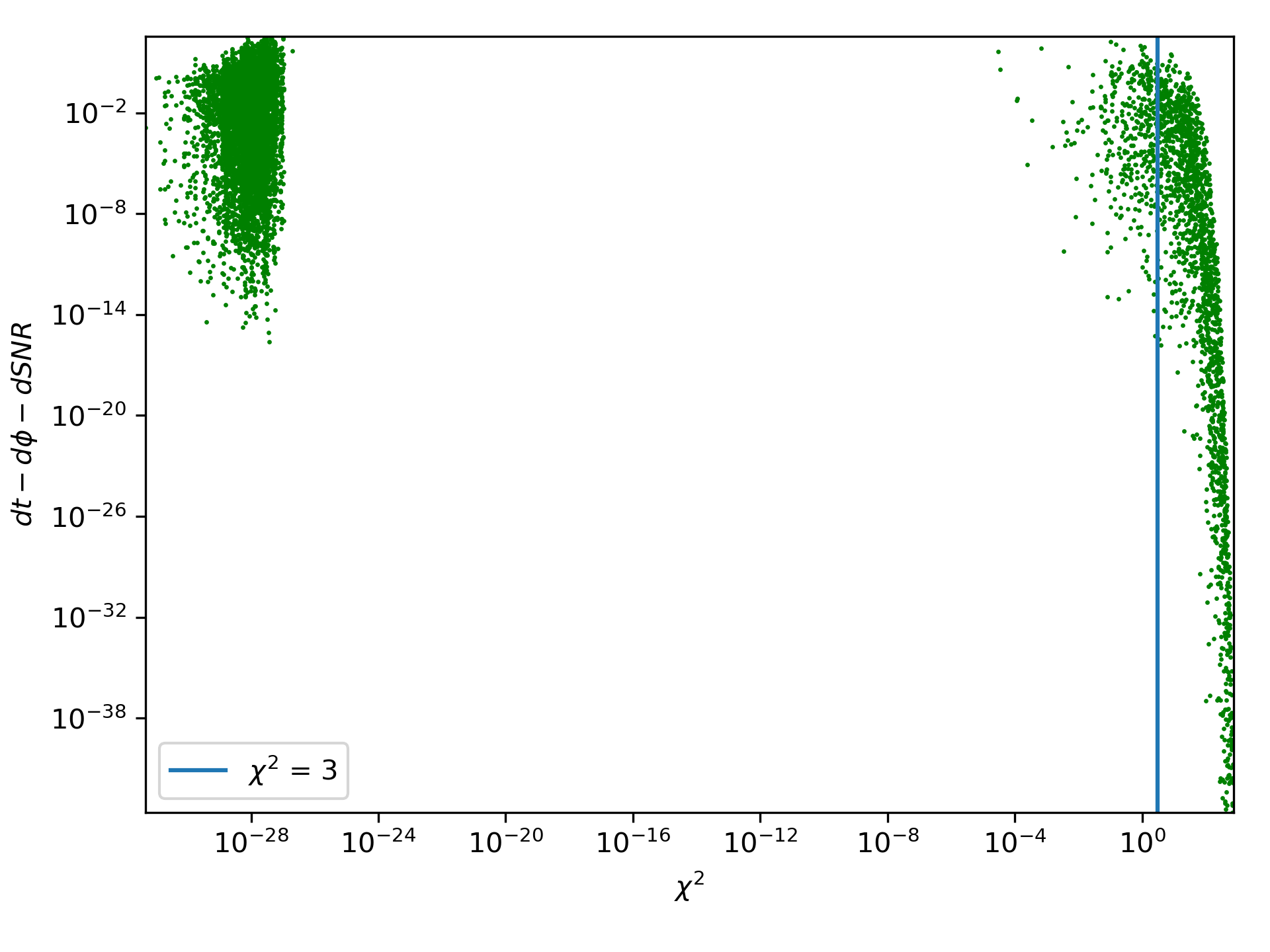}
\caption{\label{fig:dtdphi}
This figure shows the $\chi^2$ value plotted against the $dt-d\phi-dSNR$ term
of the LR implemented by GstLAL, for various triggers. The two quantities, while
not perfectly correlated, are well correlated, and selecting the $\chi^2$ = 3
threshold for calling a coincidence physical is almost equivalent to selecting
a threshold value for $dt-d\phi-dSNR$. The tiny values of $\chi^2$ seen on the 
left side of the plot are numerical noise in the calculation of $\chi^2$, and 
actually represent a value of 0.
}
\end{figure}

\begin{figure}
\includegraphics[width=\columnwidth]{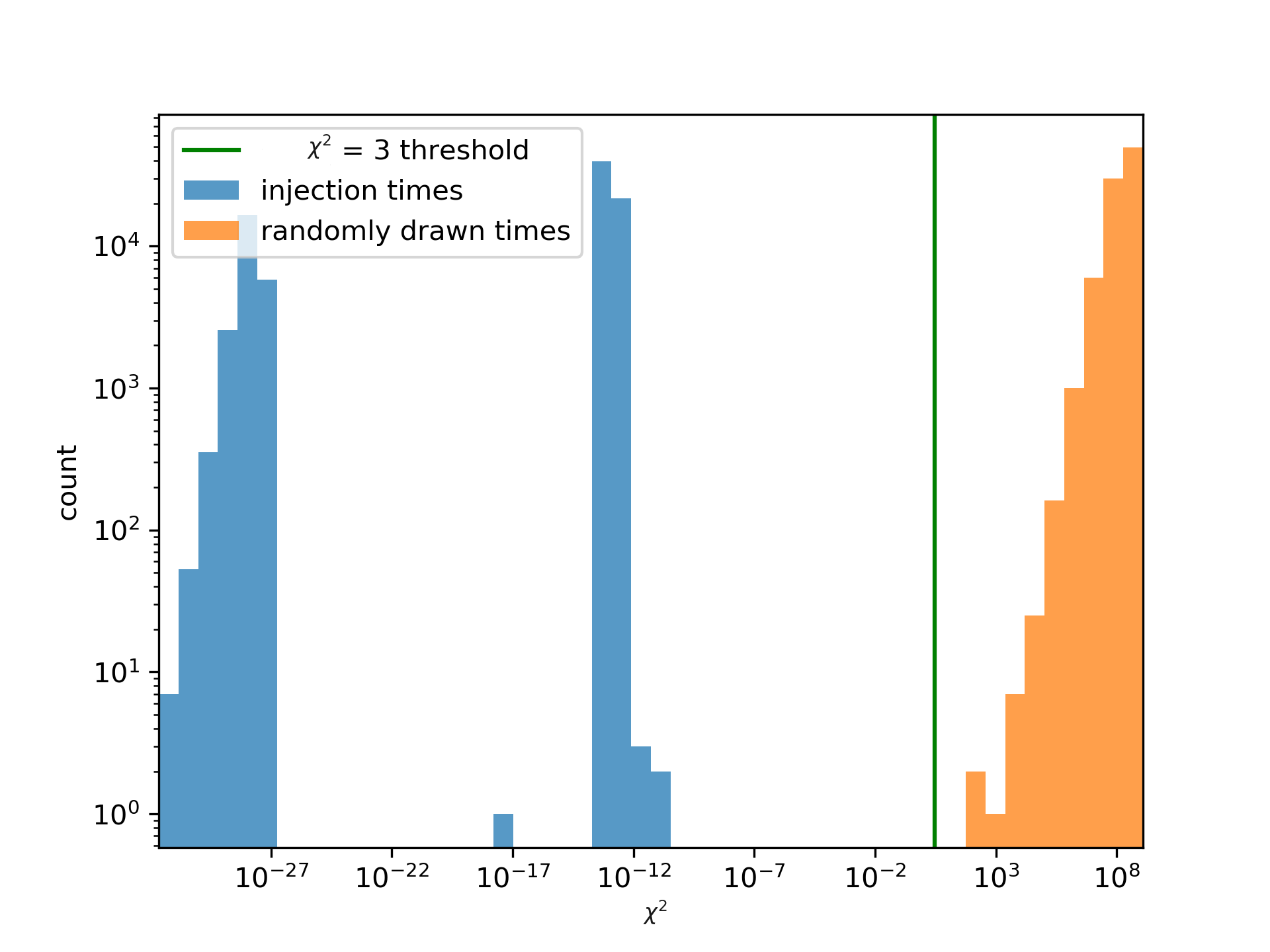}
\caption{\label{fig:chisq_inj}
This plot shows a histogram of the $\chi^2$ value for two types of triggers. The first,
represented in blue,
are simulated \ac{GW} signals, and are guaranteed to be physical. The
second, represented in orange, are triggers with randomly drawn times, and are highly
unlikely to be physical. We see that the $\chi^2$ = 3 threshold serves as a perfect
discriminator between physical and unphysical triggers. The tiny values of $\chi^2$
for the injection triggers are numerical noise in the calculation of $\chi^2$, and
actually represent a value of 0.
}
\end{figure}

Since this only considers information from the arrival times, and not phases or \acp{SNR},
this is necessary, but not sufficient for forming a physical coincidence. However, we will
show that considering time information is also sufficient. \figref{fig:dtdphi} shows that
$dt-d\phi-dSNR$ contains more information than $\chi^2$, but despite that the two are
well correlated. If we had used a threshold on $dt-d\phi-dSNR$ to qualify a trigger as 
physical instead, it would have had a very similar effect to  the $\chi^2$ = 3 threshold.
In \figref{fig:chisq_inj}, we see that the $\chi^2$ = 3 threshold is able to perfectly
distinguish between simulated gravitational wave triggers (and hence guaranteed to be
physical) and triggers generated by randomly drawing arrival times (and hence unlikely to
be physical triggers).

\subsubsection{O4 Configuration}
The SNR Optimizer has been adopted by GstLAL for use during \ac{O4}. Here,
we will provide details about the configuration used for the SNR Optimizer in \ac{O4}.
The initial template bank used by the SNR Optimizer is the full \ac{O4} GstLAL template
bank, since it has been generated via \texttt{manifold}. It is generated in the 3 dimensional
parameter space of m1-m2-$\chi_{eff}$, where $\chi_{eff}$ is the component mass weighted
average of the dimensionless spins. It contains around 2 million templates with
component masses from 1-200$M_\odot$ and dimensionless spins up to $\pm 0.99$. More
information about the \ac{O4} GstLAL template bank can be found in~\cite{Sakon:2022ibh}.

For matched filtering, the FFT length for PSD measurement and data whitening is set
to 16 seconds worth of data, and the rate of sampling the data is set to 8192 Hz.
Both of these values are larger than those used by GstLAL, which uses an FFT length
of 4 seconds of data, and a sample rate of 2048 Hz.
The threshold of standard deviations after which to gate the whitened data is set to 15.
The stopping criterion of the
minimum area of a rectangle below which it will not split into new templates is set
to the area corresponding to a mismatch of 0.001. 
This value was chosen to ensure good convergence of the search (to 99.9\%, by definition) to the maximum SNR
peak, while still maintaining reasonable latencies of less than 4 minutes.

Communication between the GstLAL low-latency analysis and the SNR Optimizer is handled
via the \texttt{Apache Kafka} software~\cite{kafka}. When a GstLAL low-latency job
finds a candidate, it sends a \texttt{kafka} message to the ``events" \texttt{kafka}
topic. From here, a specialized uploader job reads this message and uploads candidates
to GraceDB. Since there can be multiple triggers reported by different GstLAL jobs
for the same candidate, the uploader job checks at regular intervals and only uploads
the best trigger for a given candidate.
The best trigger is defined as the trigger with the lowest \ac{FAR}, if the \ac{FAR}
of the trigger is greater than the public alert threshold~\cite{opa}, and the trigger
with the highest \ac{SNR} otherwise.
After it uploads a trigger, it sends a \texttt{kafka}
message to the ``uploads" \texttt{kafka} topic. To account for multiple uploads
in a short burst of time, 10 SNR Optimizer jobs are running in parallel. Due to the 
partition structure of \texttt{kafka} topics, the message to the uploads topic
automatically gets assigned to one particular SNR Optimizer job. After that job finishes
its processing and finds a trigger for the candidate, it sends a \texttt{kafka} message
to the uploads topic, from where the uploader job can upload this trigger if it has
a higher \ac{SNR}. If it does so, SNR Optimizer jobs will not re-trigger on the 
message the uploader job will send to the uploads topic.

\section{Results}
\label{sec:results}
\subsection{Data set}
In order to test the performance of the SNR Optimizer, we set up a GstLAL
low-latency analysis, along with the SNR Optimizer, on data from \ac{O3}.
This was part of a Mock Data Challenge (MDC), and the \ac{O3} data was streamed
from Jan 02 18:39:42
UTC 2024 to Feb 11 18:39:42 UTC 2024. The MDC also
included an injection campaign. Injections are simulated
gravitational wave signals added to the data, and their purpose is to test
the performance of the analysis. More information about the MDC and the 
injection distribution used can be found in~\cite{mdc_analytics}.

During this time, the GstLAL low-latency analysis made 14710 uploads from the
injection part of the analysis. For these GstLAL candidates, the SNR Optimizer
was able to find a higher \ac{SNR} for 10259 of them, representing 70\%
of the total GstLAL uploads. Since GstLAL uploads multiple
triggers for the same candidate, after accounting for this fact, there were
5022 distinct injection candidates that GstLAL had uploads for. Out of these,
the SNR Optimizer had the highest \ac{SNR} for 2940 of them.
The trigger having the highest
\ac{SNR} for a candidate if called the ``preferred event".
In other words, the
SNR Optimizer was the preferred event for around 60\% of candidates found by GstLAL.
These injection uploads from the GstLAL low-latency analysis as well as from
the SNR Optimizer will be used for calculating the 
results presented below.

\subsection{SNR improvement}
\begin{figure}
\includegraphics[width=\columnwidth]{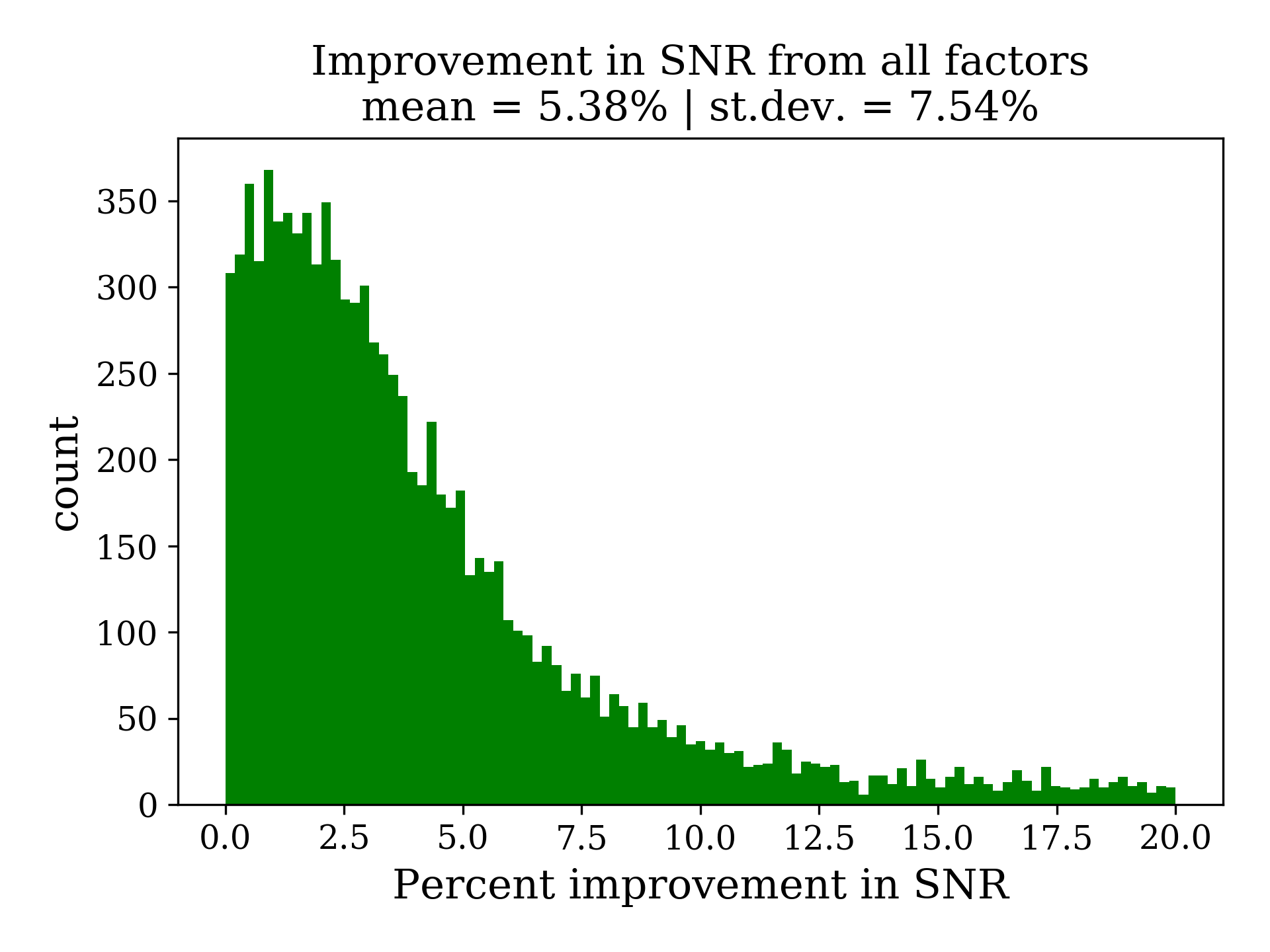}
\caption{\label{fig:snr_ratio}
A histogram of the combined \ac{SNR} improvement calculated from the
SNR Optimizer uploads as compared to GstLAL uploads.
}
\end{figure}

By comparing the network \acp{SNR} of the GstLAL uploads and the corresponding
SNR Optimizer uploads, we can create a histogram of combined \acp{SNR} improvement
due to all methods implemented by the SNR Optimizer. This histogram is shown
in \figref{fig:snr_ratio}. We see that on average, there is a 5.38\% improvement
in \ac{SNR}. A point to note here is that since the job responsible
for uploading both the GstLAL and SNR Optimizer triggers to GraceDB only does so
if the trigger has a higher \ac{SNR} than previous triggers uploaded for the 
same candidate, only the 70\% cases in which the SNR Optimizer found a higher
\ac{SNR} than GstLAL participate in this histogram.

\begin{figure*}
\includegraphics[width=\textwidth]{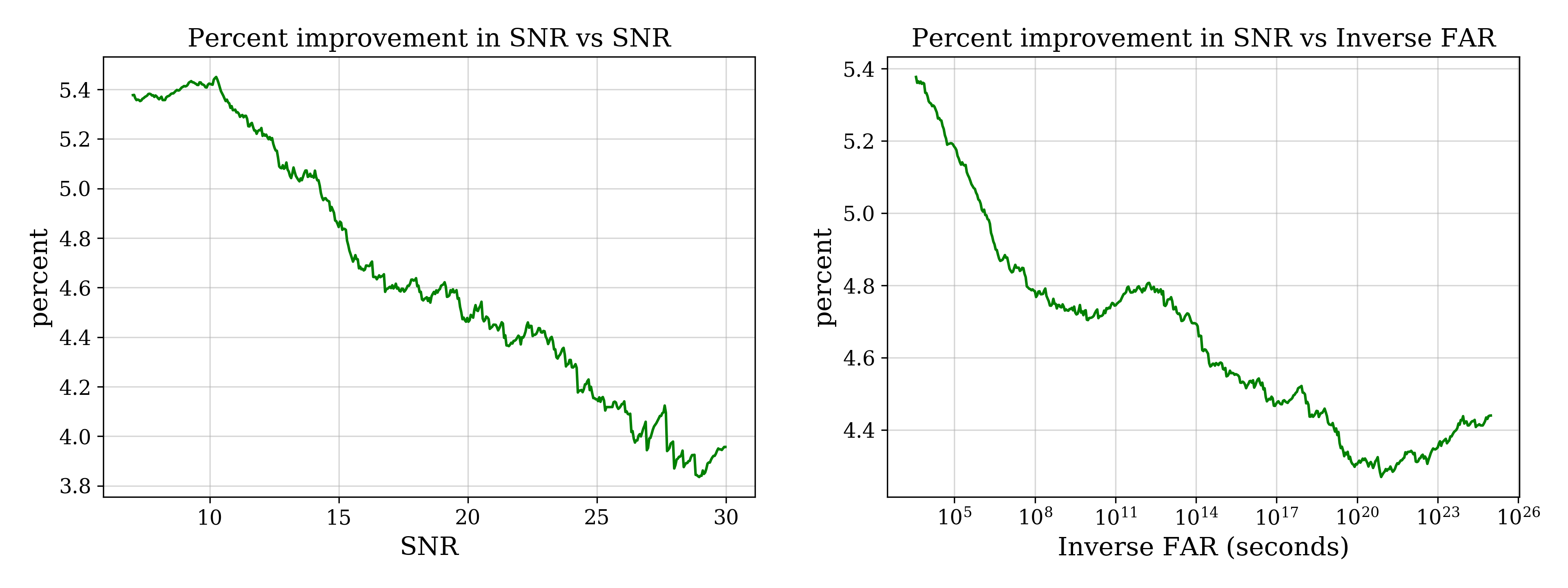}
\caption{\label{fig:imp}
This plot shows the mean percent improvement in \ac{SNR} due to the SNR Optimizer
as a function of minimum \ac{SNR} (left) and minimum inverse \ac{FAR} (right) threshold.
In both cases, the mean improvement goes down slightly with an increase in \ac{SNR} or inverse \ac{FAR}.}
\end{figure*}

\begin{figure*}
\includegraphics[width=\textwidth]{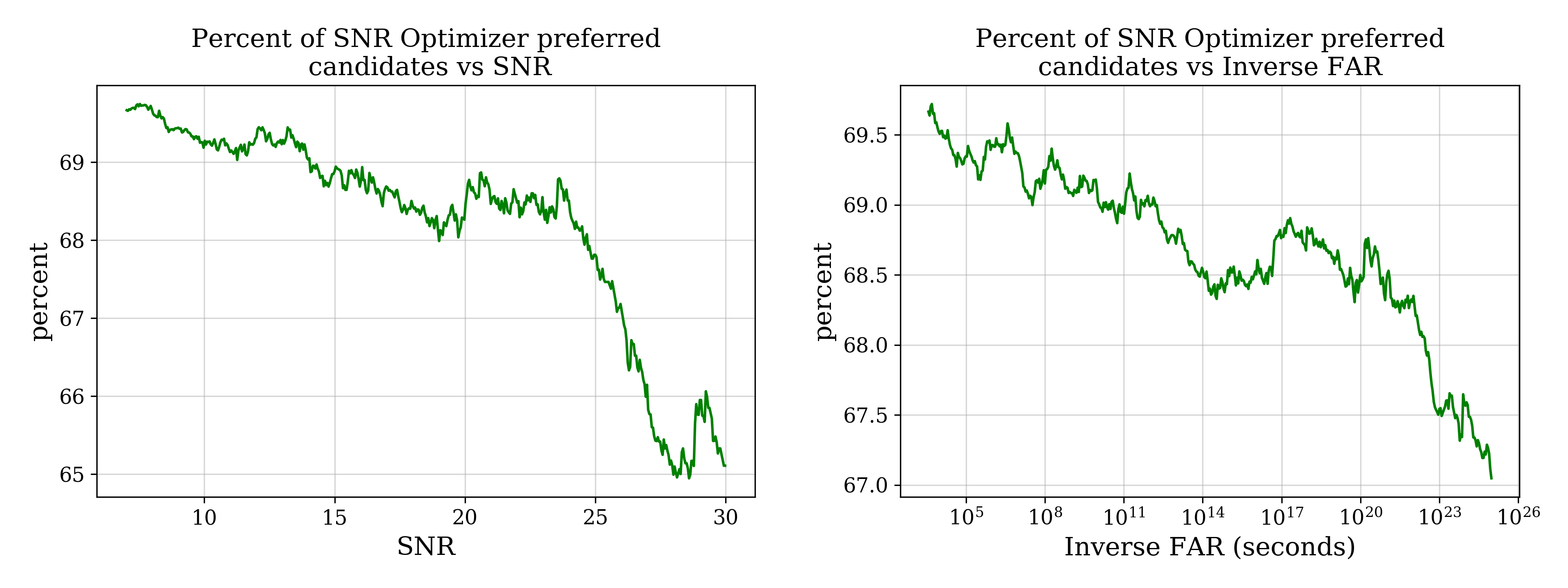}
\caption{\label{fig:num}
This plot shows the average percent of times the SNR Optimizer finds a higher SNR than GstLAL
(i.e. it is the preferred event) for a candidate, as a function of minimum \ac{SNR} (left)
and minimum inverse \ac{FAR} (right) threshold. In both cases, the average
preferred event percent goes down slightly with an increase in \ac{SNR} or inverse \ac{FAR}.}
\end{figure*}

Additionally, we can plot the mean \ac{SNR} improvement for different
\ac{SNR} and inverse \ac{FAR} thresholds. This is shown in \figref{fig:imp}.
It shows that the SNR Optimizer is most effective at lower \acp{SNR} or 
inverse \acp{FAR}, and the effectiveness goes down slightly with an increase
in \acp{SNR} or inverse \acp{FAR}.
Similarly, we can also plot the average percent of times the SNR Optimizer
is the preferred event for a candidate for different \ac{SNR} and inverse \ac{FAR} thresholds.
This is shown in \figref{fig:num}. Similar to the previous figure, we see
the same trend of the effectiveness of the SNR Optimizer going down
slightly with an increase in \acp{SNR} or inverse \acp{FAR}.

\subsection{Sky map improvement}
In this section, we will calculate the improvement in the sky map produced
by the higher \ac{SNR} of the SNR Optimizer triggers, as compared to GstLAL
triggers. The following tests are usually used to evaluate the performance
of an inference pipeline by assuming the input trigger files are correct
(see~\cite{PhysRevD.89.084060, Singer:2015ema}), but for our purposes, we
will use it to evaluate the performance of two sets of trigger files, one from
the SNR Optimizer and one from GstLAL, assuming the inference pipeline is correct.

\begin{figure}
\includegraphics[width=\columnwidth]{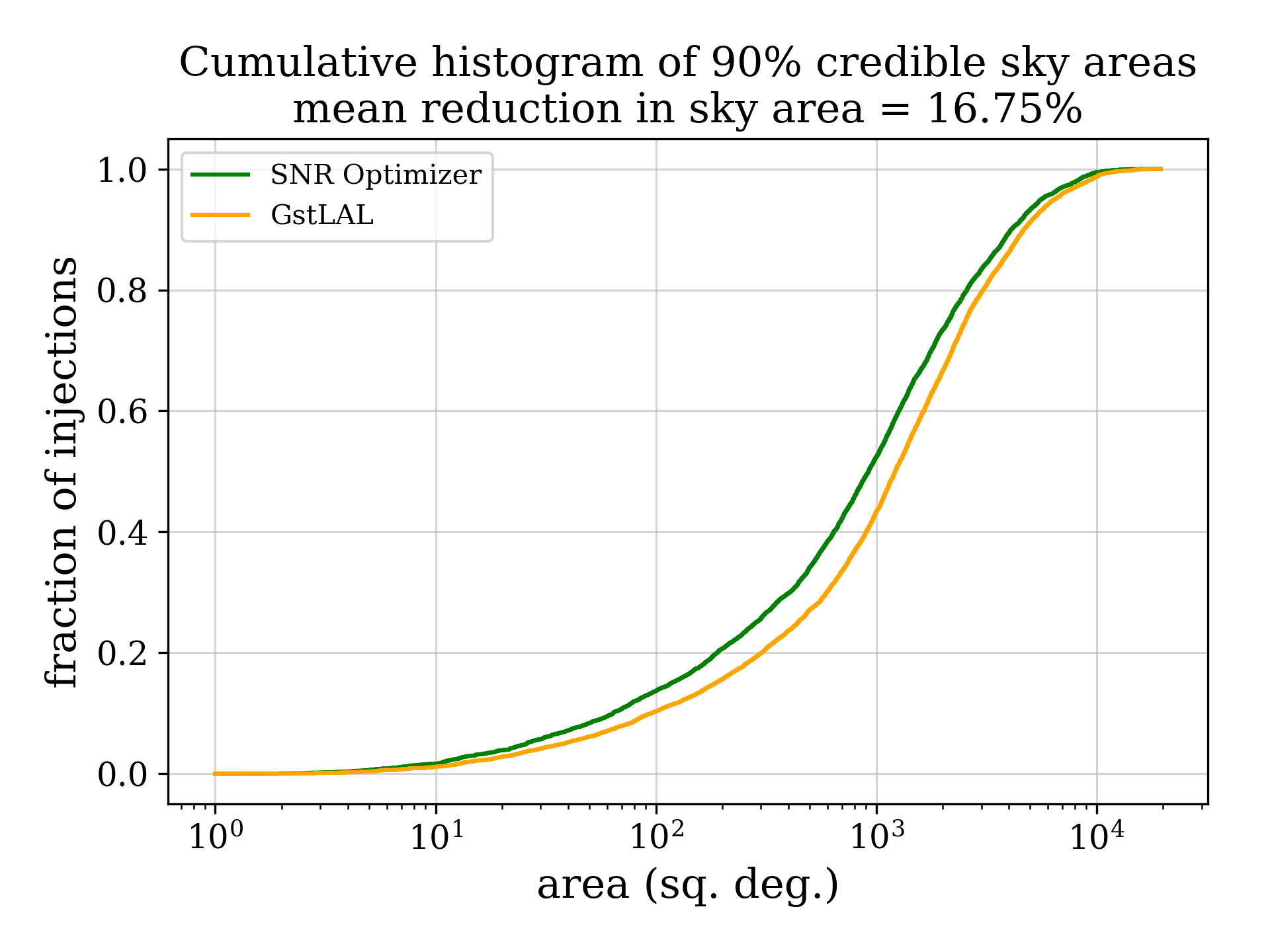}
\caption{\label{fig:sky_area}
This plot shows cumulative histograms of the 90\% sky area for
GstLAL and the SNR Optimizer. We see that sky maps produced from
the SNR Optimizer triggers are on average more constrained than those
produced from GstLAL triggers. In other words, the SNR Optimizer
results are more precise than those of GstLAL.
}
\end{figure}

In order to create a sky map from the trigger files of the SNR 
Optimizer and GstLAL, we use the BAYESTAR~\cite{Singer:2015ema, Singer:2016eax} package.
BAYESTAR, like other inference pipelines, produces a sky map that contains 
sky locations for different confidence levels. In \figref{fig:sky_area}, we compare the
sky location areas for the 90\% confidence level for both GstLAL and the 
SNR Optimizer. It shows that the sky maps produced from the SNR Optimizer
triggers are on average 16.75\% smaller than those produced from GstLAL triggers.

\begin{figure}
\includegraphics[width=\columnwidth]{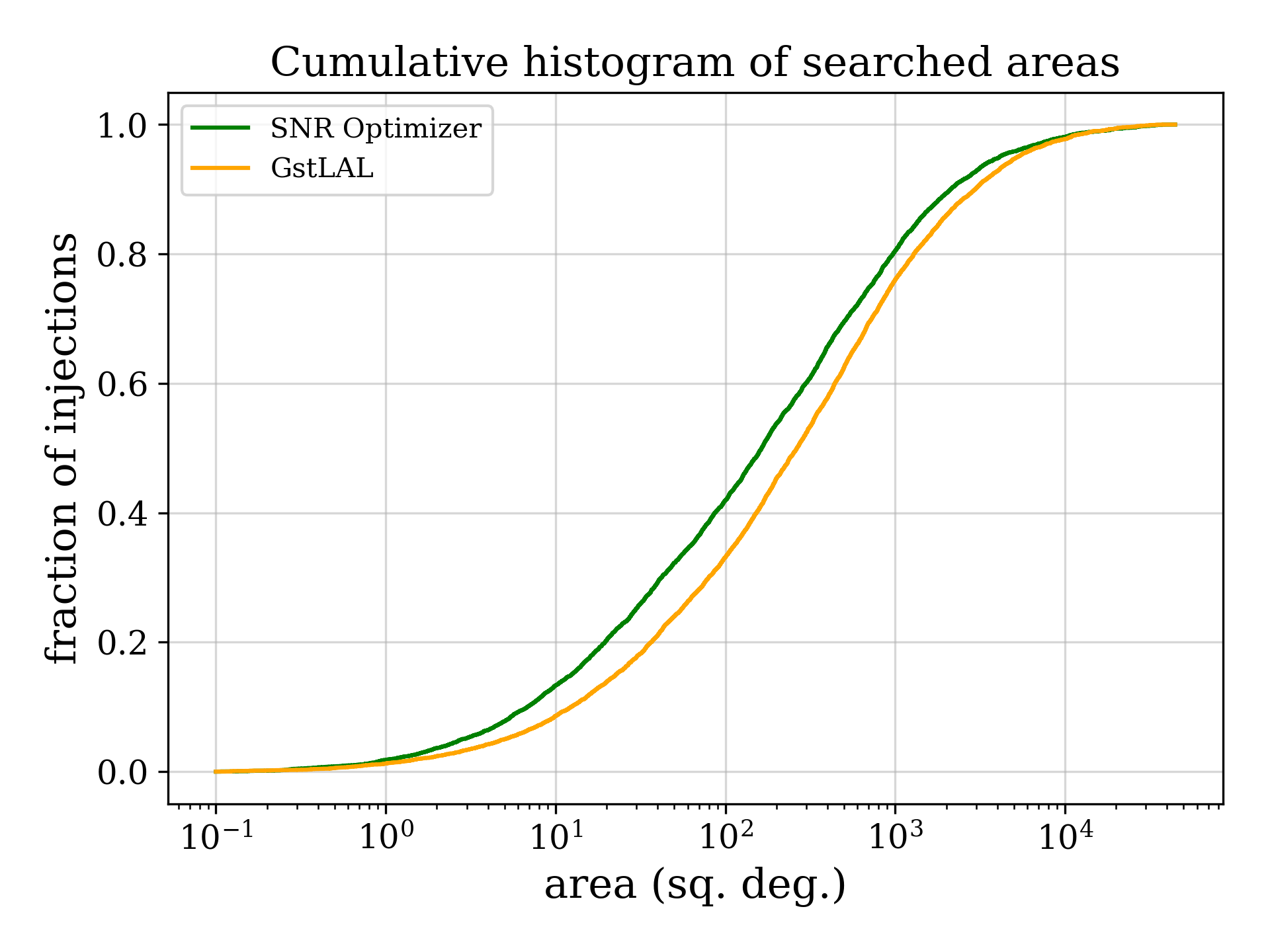}
\caption{\label{fig:searched_area}
This plot shows cumulative histograms of the searched area for
GstLAL and the SNR Optimizer. We see that sky maps produced from
the SNR Optimizer triggers on average have a  lower searched area
than those produced from GstLAL triggers. This means we have to go
to lower confidence levels for the SNR Optimizer for the sky map
to exclude the true location of the trigger, as compared to GstLAL,
implying that the SNR Optimizer results are more accurate than
those of GstLAL.
}
\end{figure}

However, a reduction in the sky area by itself only tells us that the
SNR Optimizer results are more precise, but not whether they are more accurate.
Since these results were calculated from an injection campaign,
for which we know the true sky location of all triggers, we can use that
to gauge the accuracy of the SNR Optimizer results. To do this, we
define two quantities: the searched probability, and the searched area.
The searched probability is the minimum confidence level at which the 
true location of the trigger is still within the sky map. The searched
area is the area of the sky map at that confidence level. These can be
thought of as a measure of the accuracy of the SNR Optimizer results.
The searched area results for GstLAL and the SNR Optimizer are shown in
\figref{fig:searched_area}. We see that in addition to higher precision,
the SNR Optimizer results are also more accurate than those of GstLAL.

\begin{figure}
\includegraphics[width=\columnwidth]{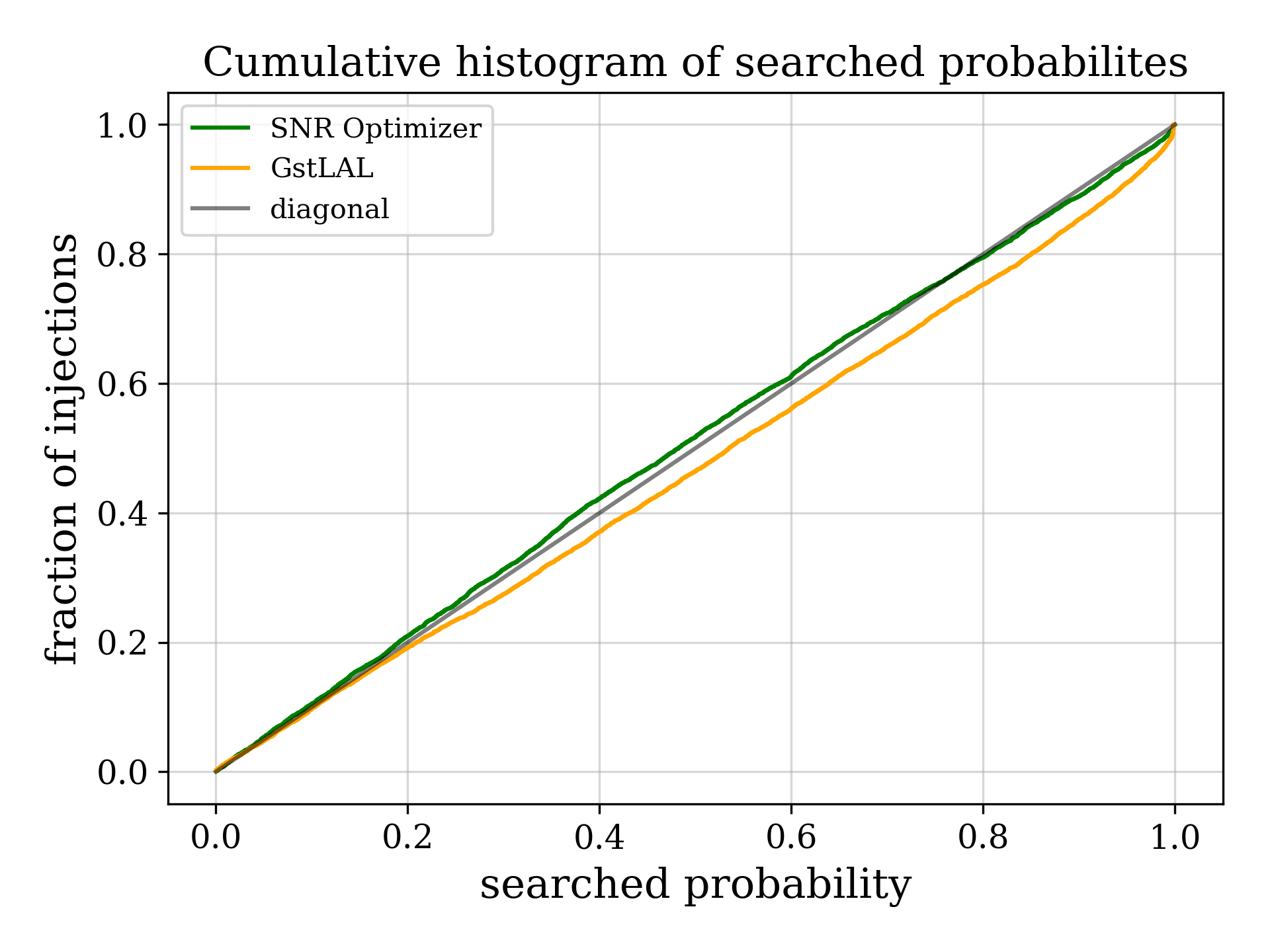}
\caption{\label{fig:searched_prob}
This plot shows cumulative histograms of the searched probability for
GstLAL and the SNR Optimizer. The SNR Optimizer line is closer to the
diagonal than the GstLAL line. This means that the higher precision and
accuracy of the SNR Optimizer is also self-consistent. This plot is
commonly called a P-P plot.
}
\end{figure}

Finally, we require a self-consistency condition relating the accuracy
and precision results presented above. We require that on average,
90\% of triggers have their true location contained within their
90\% confidence level sky map, and so on for every confidence level.
This test, commonly called a P-P plot, relates the searched probability
to the fraction of injections having that value of searched probability.
The result is shown in \figref{fig:searched_prob}, and it shows that
the SNR Optimizer results are self-consistent.

\subsection{Latency}
We can calculate two different types of latencies for the SNR Optimizer.
The first is the end-to-end latency, which is the time between the 
\ac{GW} signal reaching Earth, and the trigger
from the SNR Optimizer being uploaded to GraceDB. This includes all sources
of latency, like the data distribution latency, the latency of the initial
GstLAL trigger, the internal processing latency of the SNR Optimizer,
and any latency incurred by the uploading process.
The second type of latency is the internal processing latency recorded 
by the SNR Optimizer.

\begin{figure*}
\includegraphics[width=\textwidth]{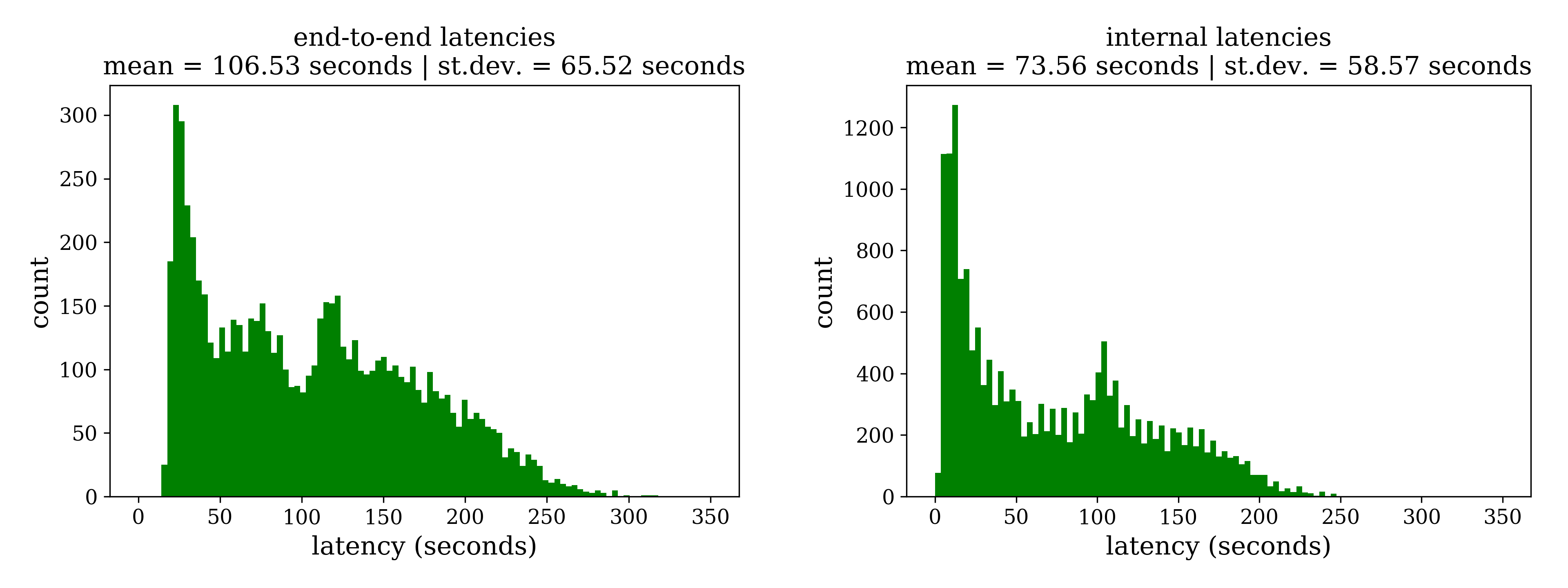}
\caption{\label{fig:latencies}
The end-to-end latencies (left) and internal latencies (right) of the SNR Optimizer. The SNR Optimizer has an internal 
timeout of 240 seconds, causing the internal latencies to be cut off
at that value, and the end-to-end latencies to only rarely exceed that.
}
\end{figure*}

A histogram of these two types of latencies,
calculated for all SNR Optimizer injection triggers from the MDC
is shown in \figref{fig:latencies}. Since the SNR Optimizer has an internal
timeout of 240 seconds, we see the histogram for the internal latencies
stop at that value.
Note that the internal latencies are recorded
by the SNR Optimizer jobs, and are not calculated from uploads to GraceDB,
and hence the dataset used for this histogram is larger than the set of 
uploads, because triggers do not get uploaded to GraceDB if they are not 
a better trigger than all triggers before.
These figures show that the typical latency
for the SNR Optimizer is around 100 seconds, and the timeout of 240 seconds
is only rarely hit.

\subsection{Contributions to the SNR improvement}
As discussed in \secref{sec:methodology}, the SNR Optimizer finds a higher \ac{SNR}
than GstLAL because of 4 main features: finding a better template, real-time
template whitening, higher FFT length, and higher data sampling rate. In order to
quantify the effects of each, we set up an offline SNR Optimizer run with all of these
features turned off, and then new runs with each feature turned on separately. By
comparing the two, we can get an estimate of the contribution of each in the
overall \ac{SNR} improvement. Note that this is a fully offline test, and hence
does not have any effects from the selective uploads performed by an online run.
As such, the results here are not directly comparable to the online results presented
in the previous subsections.

\begin{figure*}
\includegraphics[width=\textwidth]{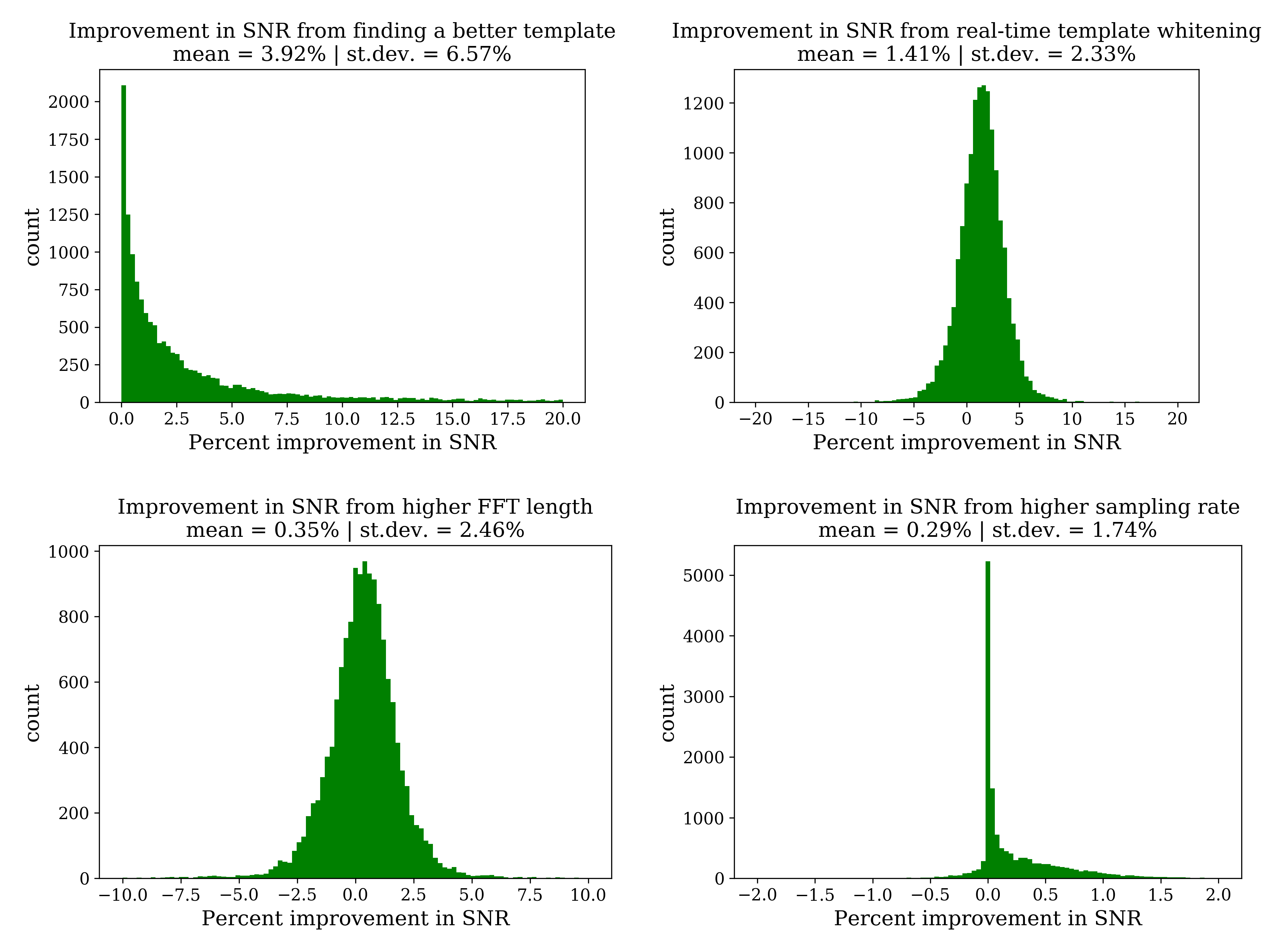}
\caption{\label{fig:cont}
This plot shows the contributions from finding a better template (top left),
real-time template whitening (top right), higher FFT length (bottom left),
and higher data sampling rate (bottom right) towards the overall SNR gain
obtained by the SNR Optimizer. During the process of finding a better template, 
the templates with the maximum SNR is selected, and hence the contribution from
that is strictly positive. All other factors are only guaranteed to produce 
positive contributions \textit{on average}. For example, since the PSD can never
be perfect due to non-stationary components in the noise, using the PSD measured
in real-time to whiten the templates will not always produce a higher SNR, but will
do so on average.
}
\end{figure*}

The result of this test is shown in \figref{fig:cont}. We see that finding a better
template has on average the highest contribution, at 3.92\%, followed by real-time
template whitening at 1.41\%, higher FFT length at 0.35\%, and higher data sampling
rate at 0.29\%.

\section{Conclusion}
In order to facilitate the multi-messenger follow-up of \ac{GW} candidates,
it is necessary to provide an accurate sky map showing the location of the 
source of the candidate, so that astronomers know where to point their telescopes.
This is done by \ac{GW} inference pipelines like BAYESTAR, which in turn relies
on low-latency \ac{GW} search pipelines like GstLAL, MBTA, PyCBC, and SPIIR, to provide information
about the arrival times, phases, and amplitudes of the signal at the different
\ac{GW} detectors.

The estimates of these provided by \ac{GW} search pipelines might
not be completely accurate due to 4 main reasons: discreteness of the template bank
used to find signals in the data, lack of template whitening using a real-time PSD,
low FFT length used to measure the PSD and whiten the data, and low sampling rate
of the data. The latter two are done to reduce latencies of the low-latency analysis.
While some of these factors are common to all \ac{GW} search pipelines, others are
specific to GstLAL, the search pipeline that this work focuses on.

We introduced the GstLAL SNR Optimizer as a way to minimize these effects,
get higher \acp{SNR} and consequently better sky maps in a low-latency
GstLAL analysis. It does this by setting up a small targeted, hierarchical,
sub-threshold search around GstLAL candidates, in a medium-latency timescale.
As long as GstLAL does not report any \ac{GW} candidates,
the SNR Optimizer keeps ingesting data, measuring its PSD, and whitening the
data with it. As soon as a candidate is reported, the SNR Optimizer
finds the 1000 nearest templates to the template reported by GstLAL, and whitens them
with the latest estimate of the PSD it has calculated. It also grabs the
the whitened data in a 2 second interval around the time reported by GstLAL, and matched filters
this data with the 1000 whitened templates. It retains the 500 highest SNR templates out
of these, and splits each of them into two new templates occupying the holes of the previous
template bank. To do this, is uses a metric defined on the intrinsic parameter space.
As this process keeps repeats, the SNR Optimizer reaches closer and closer to the
true parameters of the signal. Templates are not split if their parameter space area
is lower than a preset threshold. The process stops when no more templates can split,
or an internal timeout of 240 seconds is hit. In this way, the SNR Optimizer
is able to find the best matching template for a \ac{GW} signal. Along with
real-time template whitening, it can also afford to implement a higher FFT length
and a higher data sampling rate, leading to higher \acp{SNR}.

To test the efficacy of the SNR Optimizer, we set up a low-latency GstLAL analysis
alongside the SNR Optimizer on 40 days of \ac{O3} data. We found that the
SNR Optimizer is able to find higher \acp{SNR} roughly 70\% of the time, and that
when it does so, it improves the \ac{SNR} by 5.38\% on average. This increase
in SNR translates to a 16.75\% reduction in the size of the sky map derived from
the SNR Optimizer trigger. We also showed that such sky maps are more accurate and 
self-consistent as compared to those derived from GstLAL triggers. The SNR Optimizer
is able to do this with latencies of 100 seconds on average. Finally, we showed that
the factor contributing the most to increased \acp{SNR} is finding a better template,
followed by real-time template whitening, higher FFT length, and higher data sampling
rate.

With more electromagnetically bright sources of \acp{GW} expected to be detected in the
future, tools like the SNR Optimizer are likely to prove useful in facilitating
multi-messenger astronomy. GstLAL has already adopted the SNR Optimizer for its
low-latency operations during \ac{O4}.

\begin{acknowledgments}
This research has made use of data, software and/or web tools obtained from the
Gravitational Wave Open Science Center (https://www.gw-openscience.org/ ), a
service of \ac{LIGO} Laboratory, the \ac{LSC} and the Virgo
Collaboration.  
We especially made heavy use of the \ac{LVK} Algorithm
Library. 
\ac{LIGO} was constructed by the California Institute of Technology and the 
Massachusetts Institute of Technology with funding from the United States 
National Science Foundation (NSF) and operates under cooperative agreements 
PHYS-$0757058$ and PHY-$0823459$.
In addition, the Science and Technology Facilities Council (STFC) of the United 
Kingdom, the Max-Planck-Society (MPS), and the State of Niedersachsen/Germany 
supported the construction of \ac{aLIGO} and construction and operation of the 
GEO600 detector. 
Additional support for \ac{aLIGO} was provided by the Australian Research Council.  
Virgo is funded, through the European Gravitational Observatory (EGO), by the 
French Centre National de Recherche Scientifique (CNRS), the Italian Istituto 
Nazionale di Fisica Nucleare (INFN) and the Dutch Nikhef, with contributions by 
institutions from Belgium, Germany, Greece, Hungary, Ireland, Japan, Monaco, 
Poland, Portugal, Spain.

This material is based upon work supported by NSF's \ac{LIGO} Laboratory which is a
major facility fully funded by the National Science Foundation.
The authors are grateful for computational resources provided by 
the \ac{LIGO} Lab cluster at the \ac{LIGO} Laboratory and supported by 
PHY-$0757058$ and PHY$-0823459$, the Pennsylvania State University's Institute 
for Computational and Data Sciences gravitational-wave cluster, 
% and the University of Wisconsin Milwaukee Nemo and supported by PHY-$1626190$ and PHY-$2110594$
and supported by 
OAC-$2103662$, PHY-$2308881$, PHY-$2011865$, OAC-$2201445$, OAC-$2018299$, 
% Chad's funding: OAC-$2103662$, PHY-$2308881$, PHY-$2011865$, OAC-$2201445$, OAC-$2018299$
and PHY-$2207728$.  
% PHY-$2207728$ is what UWM people are supported by. 
CH Acknowledges generous support from the Eberly College of Science, the 
Department of Physics, the Institute for Gravitation and the Cosmos, the 
Institute for Computational and Data Sciences, and the Freed Early Career Professorship.
MWC acknowledges support from the NSF with grant numbers PHY-2308862 and PHY-2117997.
US and SS acknowledge support from NSF PHY-2409758.

\end{acknowledgments}

\appendix
\section{Compatibility with other \ac{GW} search pipelines}
While the SNR Optimizer implements a lot of the same ideas as GstLAL,
such as matched filtering in the time domain, data and template whitening
in the frequency domain, as well as using the \texttt{GStreamer} software~\cite{gstreamer}
to stream data, it does so independently of GstLAL. As such, it is designed
almost completely modularly and can in theory be used by any other \ac{GW} search pipeline.
Also possible is to have a common set of SNR Optimizer instances listening to uploads
from all \ac{GW} search pipelines.

The initial template bank required by the 
SNR Optimizer needs to be made by \texttt{manifold}, since it needs to contain
the parameter space metric information calculated by \texttt{manifold}, but this 
initial template bank does not need to be the same as the template bank used
by the search pipeline it is listening to. Currently, there are 2 minor dependencies
that the SNR Optimizer has on GstLAL, but both of these are easy to fix:
\begin{enumerate}
\item{The SNR Optimizer relies on the GstLAL job which uploads triggers to GraceDB
to process its own uploads. However, another instance of the same job can be 
easily set up by the SNR Optimizer to make its uploads independent of the GstLAL
uploader job.}
\item{The SNR Optimizer relies on the \texttt{Apache Kafka} software package to
get information from the GstLAL low-latency analysis about candidates it found.
Since \texttt{kafka} is not necessarily used by all search pipelines, this method
of communication can instead by changed to be through GraceDB. There already exists
code to continuously communicate with GraceDB about uploads sent to it, and that
can be repurposed over here. This method will increase the SNR Optimizer latencies by a small amount,
since the communication is not happening locally like in \texttt{kafka}, but this 
increase will only be of the order of a second.}
\end{enumerate}

\section{Low-Latency Mode}
Since the SNR Optimizer processes data independently of GstLAL,
it can analyze data from more detectors than what GstLAL is doing. This
can often happen when a detector like Virgo is deemed to not be sensitive
enough to be included in the GstLAL low-latency search, since data from the Virgo detector
will be used to calculate the significance of candidates, potentially leading to a
less sensitive analysis. In such situations, the SNR Optimizer can ingest
GstLAL's candidates formed from LIGO Hanford and Livingston data, and use
that to form candidates with Virgo data added to it as well. This is particularly
useful, since adding detectors to a candidate gives by far the biggest improvement
in sky maps, as discussed in \secref{sec:introduction}.

In such a use case, since the SNR Optimizer is providing a large amount
of new information to the downstream \ac{GW} inference pipelines, rather than
an incremental amount, it is necessary to reduce the SNR Optimizer's latencies to
fall in the low-latency regime.
As discussed before, the SNR Optimizer normally operates in medium-latency,
and has an average latency of 100 seconds. This is because during the hierarchical
search process, it matched filters the data with tens of thousands of
templates. In order to reduce latencies, a few things can be done:

\begin{enumerate}
\item{The SNR Optimizer can stop doing any maximization
over templates, and to only use the template reported by GstLAL to produce the
SNR Optimizer's trigger}
\item{The SNR Optimizer can re-use the Hanford and Livingston data supplied
by GstLAL, and only matched filter Virgo data and add it to the candidate, while
still observing the coincidence formation method discussed in \secref{sec:methodology}.}
\item{The SNR Optimizer can set its FFT length to be low, like GstLAL does (4 seconds worth of data)}
\item{The SNR Optimizer can set its data sampling rate to be low, like GstLAL does (2048Hz)}
\end{enumerate}

By doing so, the SNR Optimizer is able to very quickly add Virgo data to the candidates
formed by GstLAL from Hanford and Livingston data. The average internal latency of the 
SNR Optimizer in this low-latency mode is less than 1 second, with end-to-end latencies
averaging around 10 seconds. This low-latency mode of the SNR Optimizer was used alongside
the regular mode by GstLAL for part of its \ac{O4} low-latency operations.

\clearpage

\bibliography{references}

\end{document}